\documentclass[preprint,twocolumn,trackchanges]{aastex6}
\usepackage{verbatim}
\hypersetup{linkcolor={blue},citecolor={blue}}
\shorttitle{Oxygen enrichment by massive stars}
\shortauthors{SUZUKI\&MAEDA}
\begin{document}
\title{Constraining the final fates of massive stars by oxygen and iron enrichment history in the Galaxy}
\author{AKIHIRO SUZUKI\altaffilmark{1} and KEIICHI MAEDA\altaffilmark{2}}
\altaffiltext{1}{Yukawa Institute for Theoretical Physics, Kyoto University, Kitashirakawa-Oiwake-cho, Sakyo-ku, Kyoto, 606-8502, Japan}
\altaffiltext{2}{Department of Astronomy, Kyoto University, Kitashirakawa-Oiwake-cho, Sakyo-ku, Kyoto, 606-8502, Japan}
\begin{abstract}
Recent observational studies of core-collapse supernovae suggest only stars with zero-age main sequence masses smaller than $16$--$18\ M_\odot$ explode when they are red supergiants, producing type IIP supernovae. 
This may imply that more massive stars produce other types of supernovae or they simply collapse to black holes without giving rise to bright supernovae. 
This failed supernova hypothesis can lead to significantly inefficient oxygen production because oxygen abundantly produced in inner layers of massive stars with zero-age main sequence masses around $20$--$30\ M_\odot$ might not be ejected into the surrounding interstellar space. 
We first assume an unspecified population of oxygen injection events related to massive stars and obtain a model-independent constraint on how much oxygen should be released in a single event and how frequently such events should happen. 
We further carry out one-box galactic chemical enrichment calculations with different mass ranges of massive stars exploding as core-collapse supernovae. 
Our results suggest that the model assuming that all massive stars with $9$--$100\ M_\odot$ explode as core-collapse supernovae is still most appropriate in explaining the solar abundances of oxygen and iron and their enrichment history in the Galaxy. 
The oxygen mass in the Galaxy is not explained when assuming that only massive stars with zero-age main sequence masses in the range of 9--17 $M_\odot$, contribute to the galactic oxygen enrichment. 
This finding implies that a good fraction of stars more massive than $17M_\odot$ should eject their oxygen layers in either supernova explosions or some other mass loss processes. 
\end{abstract}
\keywords{supernova: general -- nuclear reactions, nucleosynthesis, abundances -- Galaxy: abundances -- Galaxy: formation}

%%%%%%%%%%%%%%%%%%%%%%%
%%%  Introduction
%%%%%%%%%%%%%%%%%%%%%%%
\section{INTRODUCTION\label{intro}}
Massive stars are believed to end their lives by producing violent explosions through the gravitational collapse of their iron cores, i.e., core-collapse supernovae (CCSNe). 
Observations of the bright emission from CCSNe can be used to test stellar evolution scenarios and probe star-forming activities in distant galaxies. 
CCSNe are classified into several categories based on their photometric and spectroscopic properties \citep{1997ARA&A..35..309F}. 
Massive stars having their hydrogen envelopes attached before the iron core-collapse produce type II SNe, whose spectra at the maximum light are characterized by prominent hydrogen features. 
In other words, they explode in the red supergiant stage. 
Type II SNe are further classified into two subcategories depending on their light curves. 
Type IIP SNe exhibit light curves with a plateau, while type IIL SNe exhibit almost linearly declining light curves. 
Massive stars are thought to progressively lose their hydrogen and helium envelopes with increasing zero-age main sequence (ZAMS) mass owing to stellar winds, leading to stripped-envelope SNe. 
Spectroscopically, they are classified as type Ib and Ic SNe. 
However, how we can relate massive stars in a particular mass range to specific categories of CCSNe and compact remnants, neutron stars or black holes, has been a long-standing problem in stellar evolution theory \citep{1988PhR...163...13N,2002RvMP...74.1015W,2003ApJ...591..288H,2004MNRAS.353...87E,2012ARA&A..50..107L}. 
In addition, binary interactions can also shed the hydrogen and helium layers of a massive star. 
The details of the binary interaction responsible for producing stripped-envelope SNe are still poorly known, making the mapping of massive stars to specific types of CCSNe even more difficult \citep[e.g.,][]{2008MNRAS.384.1109E,2011MNRAS.412.1522S}. 

One of the approaches to revealing the link between massive stars in their final evolutionary states and various types of CCSNe is constraining the rates and the relative fractions of different types of CCSNe. 
Continuous efforts have been made to observationally measure these quantities \citep[e.g.,][]{1938ApJ....88..529Z,1991ARA&A..29..363V,1997A&A...322..431C,1999A&A...351..459C,2005A&A...433..807M,2012A&A...537A.132B}. 
Modern transient survey programs, such as the Lick Observatory Supernova Search (LOSS; \citealt{2011MNRAS.412.1419L,2011MNRAS.412.1441L,2011MNRAS.412.1473L,2011MNRAS.412.1508M}; see \citealt{2017PASP..129e4201S} for a recent update), Pan-STARRS \citep{2002SPIE.4836..154K}, the Palomar Transient Factory \citep[PTF;][]{2009PASP..121.1334R,2009PASP..121.1395L}, and so on, have been playing vital roles in measuring these fundamental quantities. 
Table \ref{table:fraction} provides the relative fractions of different types of CCSNe reported by \cite{2009MNRAS.395.1409S}, \cite{2010ApJ...721..777A}, and \cite{2011MNRAS.412.1522S}.

%%%%%%%%%%%%%%%%%%%%%%%%%%%%%%
\begin{table*}
\begin{center}
  \caption{Relative fractions ($\%$) of different types of core-collapse supernovae}
\begin{tabular}{ccccc}
\hline\hline
%Types&\cite{1999A&A...351..459C}&\cite{2009MNRAS.395.1409S}&\cite{2011MNRAS.412.1522S}\\
Types&Smartt et al.(2009)&Arcavi et al. (2010)&Smith et al.(2011)\\
\hline
II&$-$&$78$&$-$\\
IIP&$58.7$&$-$&$42.8$\\
IIL&$2.7$&$-$&$6.4$\\
IIn&$3.8$&$-$&$8.8$\\
IIb&$5.4$&$4$&$10.6$\\
Ib&$9.8$&$4$&$7.1$\\
Ic&$19.6$&$13$&$14.9$\\
Ic-BL or Ibc-pec&$-$&$2$&$4.0$\\
\hline\hline
\end{tabular}
\label{table:fraction}
\end{center}
\end{table*}
%%%%%%%%%%%%%%%%%%%%%%%%%%%%%

Results of recent observational studies aiming at revealing progenitors of a specific category of CCSNe have offered interesting findings on the evolutions and the final fates of massive stars. 
Particularly, identifying progenitors of nearby CCSNe in their pre-supernova images from archival data \citep{2003PASP..115.1289V,2004Sci...303..499S,2005PASP..117..121L,2005MNRAS.364L..33M,2006ApJ...641.1060L} has had a strong impact on this important issue. 
Direct progenitor detections of type IIP SNe have confirmed the prediction from stellar evolution theories that red supergiants are progenitors of type IIP SNe. 
However, these results posed a challenge at the same time, i.e., the so-called red supergiant problem. 
Current studies suggest that no type IIP SN progenitor is found above a threshold luminosity of $10^{5.1}\ L_\odot$ \citep[see, e.g.,][]{2009ARA&A..47...63S,2009MNRAS.395.1409S,2015PASA...32...16S}. 
The luminosity indicates a threshold ZAMS mass of $16$--$18\ M_\odot$, depending on the employed stellar evolution model \citep[see also][]{2017arXiv170906116D}. 
In other words, there is no direct evidence that massive stars with ZAMS masses above $16$--$18\ M_\odot$ explode as type IIP SNe. 

Another constraint on massive star-supernova connection is obtained by statistical approaches. 
\cite{2011ApJ...738..154H} claimed that nearby CCSNe rates measured by several survey programs are smaller than the expected value from cosmic star formation rates by a factor of $\sim 2$. 
Although there are several possibilities to explain this discrepancy between the observed CCSNe rates and star formation rates, such as dust-obscured CCSNe (\citealt{2012ApJ...756..111M,2012ApJ...757...70D}; see also \citealt{2014ARA&A..52..415M}), one simple interpretation is that a fraction of massive stars just collapse without giving rise to bright supernovae. 

A direct observational consequence of a massive star not exploding as a CCSN is sudden disappearance of a red supergiant or underluminous transients caused by the envelope ejection from a collapsing red supergiant \citep{1980Ap&SS..69..115N,2013ApJ...769..109L,2017ApJ...845..103L}. 
\cite{2008ApJ...684.1336K} have conducted monitoring observations of about one million red supergiants by using the Large Binocular Telescope (LBT) to search for such suddenly fading phenomena and found one possible candidate \citep{2015MNRAS.450.3289G}. 
By conducting follow-up observations of the candidate, \cite{2017MNRAS.468.4968A} concluded that the most likely explanation for the event was the disappearance of a red supergiant with ZAMS mass of $\sim 25\ M_\odot$. 
\cite{2017MNRAS.469.1445A} calculated the relative fraction of such events to the total number of the core-collapse of massive stars to be $0.14^{+0.33}_{-0.10}$ (at 90\% confidence) based on their survey. 
Another group \citep{2015MNRAS.453.2885R} searched for similar phenomena in the Hubble Space Telescope archival data and found one candidate, which was consistent with a $25$--$30\ M_\odot$ yellow giant. 

From a theoretical point of view, exploring the conditions to make CCSNe has a long history \citep[see, e.g.,][for reviews]{1990RvMP...62..801B,2007PhR...442...38J,2012ARNPS..62..407J,2013RvMP...85..245B}. 
Although reproducing a CCSN with a canonical explosion energy of $10^{51}$ erg is still unsuccessful, several recent theoretical studies have started quantifying the ``explodability'' of massive stars, which indicates how easily massive stars with different internal structures explode as CCSNe. 
\cite{2011ApJ...730...70O,2013ApJ...762..126O} introduced the so-called ``compactness'' parameter, which is proportional to a mass in the innermost part of a star divided by the radius within which the mass is contained. 
By using their numerical simulations, they suggested that the compactness $\xi_{2.5}$ for inner $2.5\ M_\odot$ at the time of the core bounce could be used to judge how easily massive stars can explode, which was also tested and confirmed by several independent groups \citep{2012ApJ...757...69U,2015PASJ...67..107N}. 
Despite some differences among these studies, they have reached the consensus that massive stars with $20$--$25\ M_\odot$ are relatively difficult to explode as CCSNe. 
This finding may partially agree with the observational indications that no type IIP progenitor more massive than $16$--$18\ M_\odot$ has been detected.

In this study, we consider the problem of mapping massive stars to different types of CCSNe or collapses without any CCSNe, in the light of heavy element production in massive stars and metal enrichment history of the Galaxy. 
A similar study has been done by \cite{2013ApJ...769...99B}, who investigated how limited mass ranges for CCSNe affect the reproduction of the solar isotopic abundance. 
They pointed out that assuming smaller maximum masses for CCSNe ($\leq 20 M_\odot$) progressively reduces the average oxygen mass ejected by a single CCSN, leading to significantly inefficient oxygen production. 
\cite{2017MNRAS.464..985G} also introduced a high-mass cut-off, above which no CCSN occurs, in their chemical evolution models for dwarf irregular galaxies and reached similar conclusions. 
The oxygen deficit is a natural consequence because oxygen is predominantly produced by stars more massive than $\sim20\ M_\odot$, as early studies on galactic chemical evolution have recognized \citep[e.g.,][]{1995ApJS...98..617T}. 

One possible solution to this oxygen deficit problem is that more massive stars (ZAMS masses larger than 30 $M_\odot$) indeed explode as different types of CCSNe, as suggested by \cite{2015PASA...32...16S}. 
These requirements from galactic chemical evolution potentially serve as strong constraints on our understanding of how massive stars end their lives. 
In order to obtain more detailed constraints on the mass range of CCSNe, we use a series of one-box chemical evolution models in this study. 
In addition to introducing a cut-off mass at $17\ M_\odot$ as previous studies \citep{2013ApJ...769...99B,2017MNRAS.464..985G}, we also allow more massive stars to explode. 
This is in line with recent theoretical investigations on the explodability of massive stars. 
By systematically investigating the likely mass range of massive stars ending up as CCSNe and their relative fractions, we constrain the final evolutionary state of massive stars, various types of SNe, failed SNe, or other unspecified transients, in the light of chemical enrichment history of the Galaxy. 

This paper is structured as follows. 
In Section \ref{sec:oxygen_production}, we review and summarize the current understanding of oxygen production in CCSNe. 
In order to obtain a model-independent requirement for the oxygen enrichment, we first assume an unspecified population responsible for providing the interstellar gas with oxygen. 
We consider the requirement for the population in Section \ref{sec:missing}. 
Furthermore, we use our one-box chemical evolution model to show how limited mass ranges for CCSNe affect the chemical enrichment in the Galaxy. 
Results are presented in Section \ref{sec:chemical_evolution}. 
We discuss the implications on the final fates of massive stars in Section \ref{sec:discussion}. 
Finally, Section \ref{sec:summary} summarizes this paper. 
In this paper, we adopt the solar abundance reported by \cite{2009ARA&A..47..481A}.
%, whose oxygen abundance is less than that in \cite{1989GeCoA..53..197A} by $\sim 0.3$ dex. 
%%%%%%%%%%%%%%%%%%%%%%%
%%%  Supernova yield
%%%%%%%%%%%%%%%%%%%%%%%
\section{Massive star evolution and supernova yield}\label{sec:oxygen_production}

\subsection{Initial Mass Function}\label{sec:IMF}
We introduce the stellar initial mass function  (IMF), which have extensively been investigated by many authors \citep[e.g.,][]{1955ApJ...121..161S,1986FCPh...11....1S,2001MNRAS.322..231K,2003PASP..115..763C}. 
We assume a Kroupa IMF \citep{2001MNRAS.322..231K}, which is expressed by the broken power-law function $\Phi(M)\propto M^{-\alpha}$ with
\begin{equation}
\alpha=
\left\{
\begin{array}{ccl}
0.3&\mathrm{for}&M/M_\odot<0.08,\\
1.3&\mathrm{for}&0.08\leq M/M_\odot<0.5,\\
2.3&\mathrm{for}&0.5\leq M/M_\odot.
\end{array}
\right.
\label{eq:Kroupa}
\end{equation}
We normalize the IMF so that it gives the number of stars with mass $[M,M+dM]$ per $1M_\odot$ of star formation. 
Thus, the distribution $dN/dM$ is expressed as follows,
\begin{equation}
\frac{dN}{dM}=C\Phi(M),
\label{eq:IMF}
\end{equation}
where
\begin{equation}
C^{-1}=\int^{M_\mathrm{max}}_{M_\mathrm{min}}M\Phi(M)dM.
\end{equation}
The lower and upper bounds, $M_\mathrm{min}$ and $M_\mathrm{max}$, of the integration are fixed to be $M_\mathrm{min}=0.08\ M_\odot$, which is the hydrogen burning limit \cite[e.g.,][]{2012sse..book.....K}, and $M_\mathrm{max}=100\ M_\odot$. 

\subsection{Theoretical CCSNe Yield}\label{sec:CCSNe_yield}
%%%%%%%%%%%%%%%%%%%%%%%%%%%%%
\begin{figure*}
\begin{center}
\includegraphics[scale=1.2,bb=0 0 357 223]{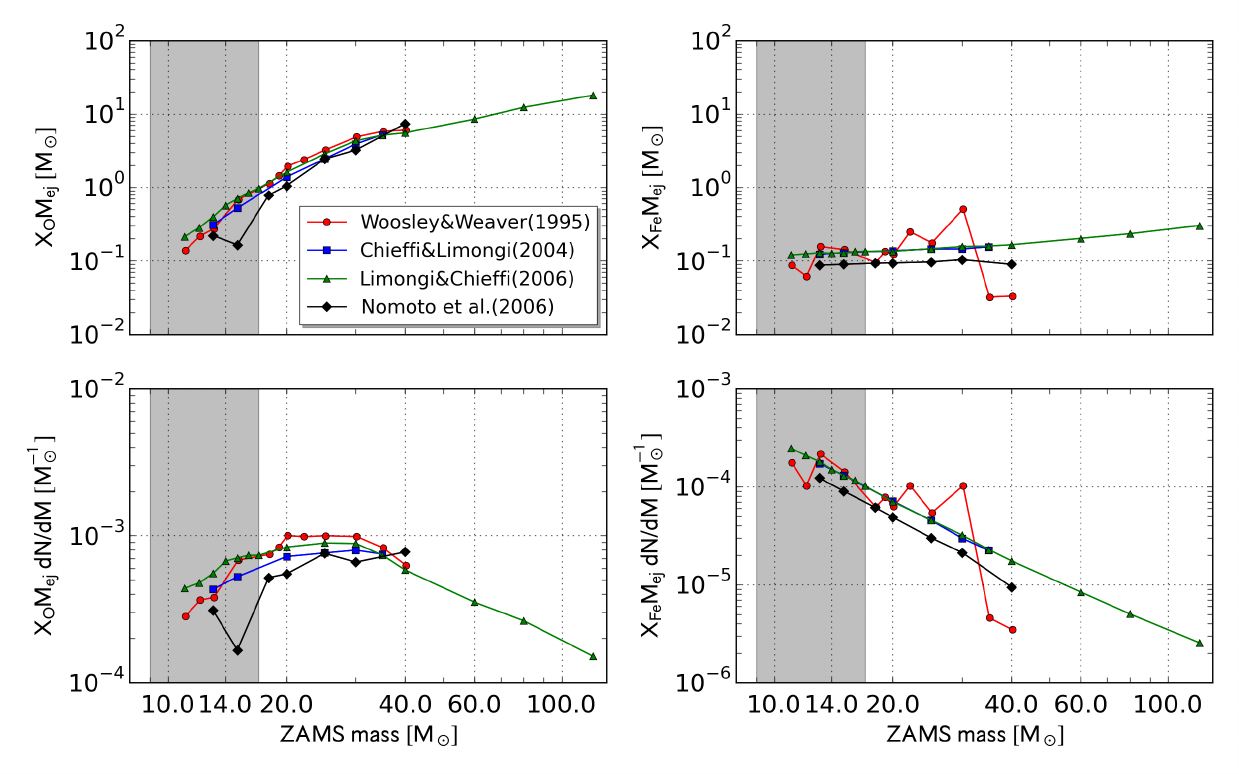}
\caption{
Oxygen (left) and iron (right) production in massive stars with various ZAMS masses. 
The upper panels show the ejected mass as a function of ZAMS mass, while the lower panels show the product of the ejecta mass and the IMF, Equation (\ref{eq:IMF}). 
Models provided by different groups are plotted in each panel (circle: \protect\cite{1995ApJS..101..181W}, square: \protect\cite{2004ApJ...608..405C}, triangle: \protect\cite{2006ApJ...647..483L}, diamond: \protect\cite{2006NuPhA.777..424N})
The mass range of $9$--$17\ M_\odot$, which corresponds to type IIP SN progenitors implied by observations \protect\citep{2015PASA...32...16S}, is represented by the shaded area. 
The models with metallicity $Z=0.02$ are plotted for \protect\cite{1995ApJS..101..181W}, \protect\cite{2004ApJ...608..405C}, and \protect\cite{2006NuPhA.777..424N}, while \protect\cite{2006ApJ...647..483L} provide zero-metallicity models. 
}
\label{fig:mass_OFe}
\end{center}
\end{figure*}
%%%%%%%%%%%%%%%%%%%%%%%%%%%%%
%%%%%%%%%%%%%%%%%%%%%%%%%%%%%
\begin{figure*}
\begin{center}
\includegraphics[scale=1.2,bb=0 0 357 223]{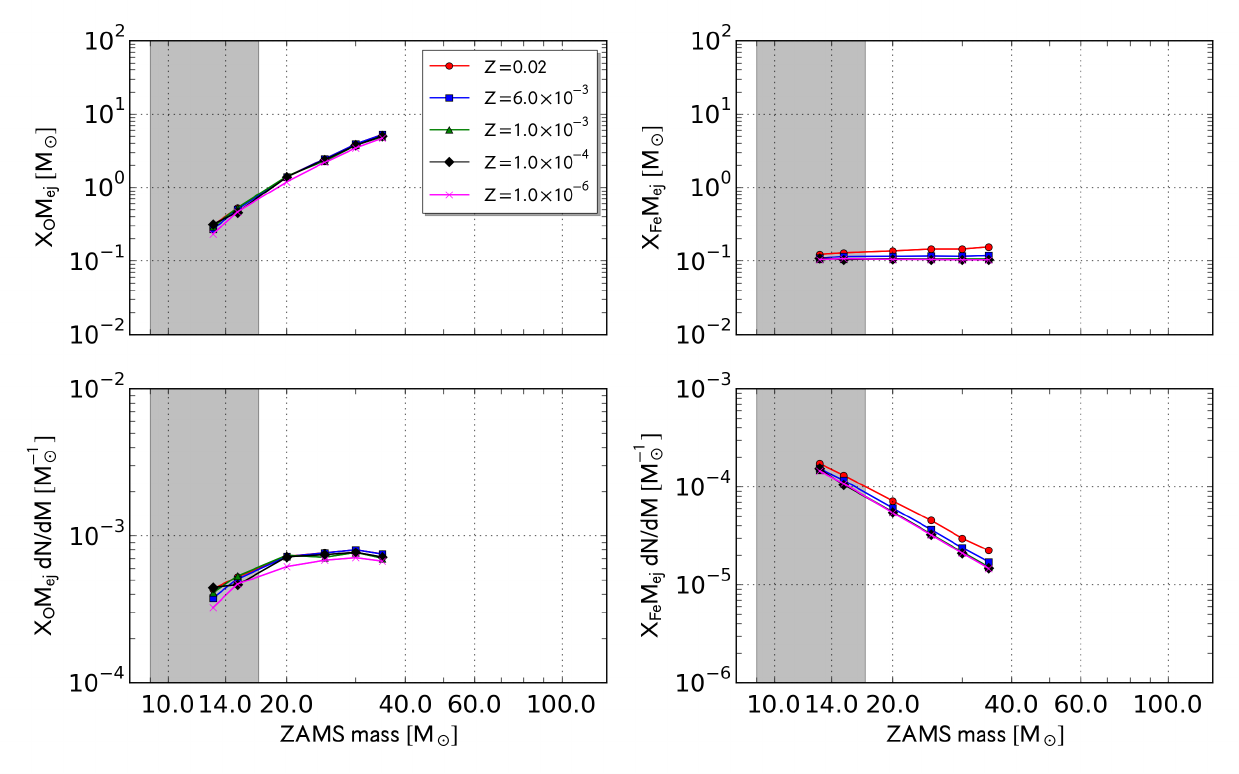}
\caption{Same as Figure \ref{fig:mass_OFe}, but for models ($M=13,\ 15,\ 20,\ 25,\ 30,$ and $35\ M_\odot$) with different metallicities calculated by \protect\cite{2004ApJ...608..405C}. 
The curves in the panels correspond to $Z=0.02$ (red circle), $6.0\times10^{-3}$ (blue square), $1.0\times10^{-3}$ (green triangle), $1.0\times10^{-4}$ (black diamond), and $1.0\times10^{-6}$ (magenta cross). 
}
\label{fig:mass_OFeZ}
\end{center}
\end{figure*}
%%%%%%%%%%%%%%%%%%%%%%%%%%%%%

Several groups have theoretically calculated evolutions of massive stars with various ZAMS masses and metallicities and provided the ejecta masses, remnant masses, and mass fractions of chemical elements produced by nuclear burnings during the stellar evolution and SN explosions. 
We denote the ejecta mass and the mass fraction of an element $i$ for given ZAMS mass $M$ and metallicity $Z$ by $M_\mathrm{ej}^\mathrm{cc}(M,Z)$ and $X_i^\mathrm{cc}(M,Z)$. 
Thus, the mass of an element $i$ in the SN ejecta is given by the product $X_i^\mathrm{cc}(M,Z)M_\mathrm{ej}^\mathrm{cc}(M,Z)$. 
We plot the masses of oxygen and iron in the ejecta provided by several groups and the product of the ejected masses and the IMF ($dN/dM$) in Figure \ref{fig:mass_OFe}. 
The top left panel of Figure \ref{fig:mass_OFe} clearly shows that the oxygen mass in the ejecta is a growing function of the ZAMS mass. 
On the other hand, the iron mass is insensitive to the ZAMS mass. 
In these models, the masses of iron-peak elements in the ejecta highly depend on the so-called ``mass cut'', a threshold mass coordinate below which the stellar mantle ends up as a compact object. 
The mass cut is artificially determined so that the mass of $^{56}$Ni is equal to a certain value (usually $\sim 0.1\ M_\odot$). 
Therefore, the iron mass as a function of the ZAMS mass in the top right panel of Figure \ref{fig:mass_OFe} is a consequence of the artificial calibration. 
Nevertheless, similar values of $^{56}$Ni mass, $\sim 0.1\ M_\odot$, indicated by bolometric light curves of various types of CCSNe suggest that similar masses of iron-peak elements are ejected by CCSNe with different ZAMS masses. 

The oxygen and iron masses multiplied by the IMF (lower panels of Figure \ref{fig:mass_OFe}) demonstrate a clear difference between the two elements. 
The contribution to the oxygen production as a function of the ZAMS mass has a peak around $20$-$30\ M_\odot$, while that to iron is generally a decreasing function of the ZAMS mass. 
Therefore, oxygen is predominantly produced by stars with ZAMS masses in the range of $20$--$30\ M_\odot$, while the dominant producers of iron are less massive stars with ZAMS masses close to the minimum mass of CCSNe (with the delayed contribution from SNe Ia). 
Early studies of supernova nucleosynthesis and chemical enrichment of galaxies have clarified the importance of stars with $20$--$30\ M_\odot$ in producing oxygen in the universe \citep{1995ApJS...98..617T}. 
This difference has a striking impact on the production of these two elements when only less massive stars can successfully explode as implied by recent progenitor searches. 
The shaded region in each panel of Figure \ref{fig:mass_OFe} represents the mass range of likely type IIP SNe progenitors suggested by \cite{2015PASA...32...16S}. 
The implied upper mass $\sim 17\ M_\odot$ is smaller than the masses of stars predominantly contributing oxygen production. 
On the other hand, since the iron distribution in the lower-right panel of Figure \ref{fig:mass_OFe} is a decreasing function of ZAMS mass, these stars can successfully produce iron. 
Therefore, although type IIP progenitors can contribute to iron production, they cannot to oxygen. 
This conclusion has also been reached by \cite{2013ApJ...769...99B}, who investigated whether limited mass ranges of type IIP SN progenitors could reproduce the solar isotopic abundance. 

Figure \ref{fig:mass_OFeZ} shows the metallicity dependence of the oxygen and iron masses. 
The models with $Z=0.02$, $6.0\times10^{-3}$, $1.0\times10^{-3}$, $1.0\times10^{-4}$, and $1.0\times10^{-6}$ provided by \cite{2004ApJ...608..405C} are plotted. 
The mass of $^{56}$Ni in the ejecta is again assumed to be $0.1M_\odot$ for all the models. 
Although the curves in the panels of Figure \ref{fig:mass_OFeZ} are slightly different from each other, they generally show similar values and trends. 

\subsection{IMF-weighted Yields}\label{sec:IMF_weighted_yield}
In order to quantitatively evaluate how different mass ranges for successful CCSNe affect the production of different elements, we define the following IMF-averaged mass of an element $i$ produced in the form of CCSNe ejecta per unit mass $(1M_\odot)$ of star formation,
\begin{equation}
Y_i^\mathrm{cc}=\int_{M_l}^{M_u}\frac{dN}{dM}M_\mathrm{ej}^\mathrm{cc}(M,Z)X_i^\mathrm{cc}(M,Z)dM.
\end{equation}
The lower and upper bounds, $M_\mathrm{l}$ and $M_\mathrm{u}$, of the integration give the minimum and maximum ZAMS masses of massive stars supposed to explode as CCSNe. 
The ejected mass by CCSNe per unit mass of star formation is obtained by summing the yields of different elements,
\begin{equation}
Y^\mathrm{cc}=\sum_iY_i^\mathrm{cc}.
\end{equation}
These yields usually depend on the metallicity $Z$ of the gas from which stars form.

%%%%%%%%%%%%%%%%%%%%%%%%%%%%%
\begin{figure*}
\begin{center}
\includegraphics[scale=0.6,bb=0 0 850 339]{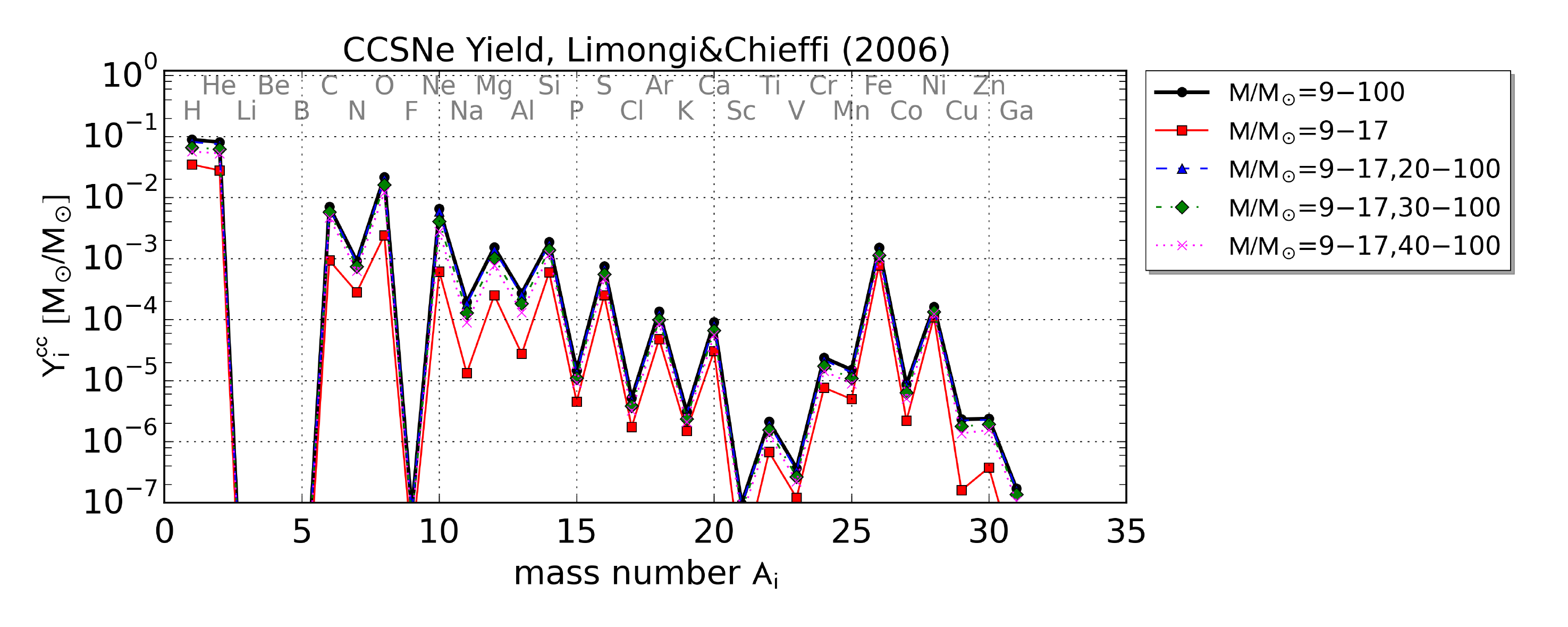}
\caption{IMF-weighted CCSNe yield of $Z=0.02$ models calculated by \protect\cite{2006ApJ...647..483L}. 
The thick solid (black) line shows the yield corresponding to the mass range of $9\leq M/M_\odot\leq 100$. 
The yields shown by filled squares (red) assume the mass range for type IIP SN progenitors, $9\leq M/M_\odot\leq 17$ \protect\citep{2015PASA...32...16S}. 
The triangles (blue), diamonds (green), and crosses (magenta) show yields obtained by assuming type IIP SN progenitors and additional contributions from more massive stars, $20\leq M/M_\odot\leq 100$, $30\leq M/M_\odot\leq 100$, and $40\leq M/M_\odot\leq 100$, respectively. 
}
\label{fig:CCSNe_yield}
\end{center}
\end{figure*}
%%%%%%%%%%%%%%%%%%%%%%%%%%%%%

In Figure \ref{fig:CCSNe_yield}, we plot the yield $Y^\mathrm{cc}_i$ as a function of the mass number $A_i$ of element $i$. 
The lines in the figure shows the IMF-weighted yield calculated by averaging over several mass ranges. 
The thick solid line represents the yields obtained by assuming that massive stars with ZAMS masses in the range of $9\leq M/M_\odot\leq 100$ successfully explode. 
When we only consider the contribution from type IIP SNe, the mass range reduces to $9\leq M/M_\odot\leq 17$, as suggested by \cite{2015PASA...32...16S}. 
The corresponding yields are plotted as the thin solid line. 
This shows clear discrepancies between the yields corresponding to the two different mass ranges. 
The oxygen yield for $9\leq M/M_\odot\leq 17$ is smaller than that for $9\leq M/M_\odot\leq 100$ by an order of magnitude.

In addition to type IIP SNe, contributions from more massive stars may be expected. 
The yields with additional contributions from stars with $20\leq M/M_\odot\leq 100$, $30\leq M/M_\odot\leq 100$, and $40\leq M/M_\odot\leq 100$ are plotted in Figure \ref{fig:CCSNe_yield}. 
In other words, these models assume that only stars with $17\leq M/M_\odot\leq 20$, $17\leq M/M_\odot\leq 30$, and $17\leq M/M_\odot\leq 40$ do not explode as CCSNe. 
Adding the increasingly larger contributions from most massive stars result in chemical yields more similar to that for $9\leq M/M_\odot\leq 100$. 

\subsection{Observational Constraints on Oxygen and Nickel Masses}\label{sec:obs_cons}
The oxygen and iron masses ejected by a CCSN can be constrained by photometric and spectroscopic observations of individual CCSNe. 
The mass of $^{56}$Ni, which successively decays to $^{56}$Co and $^{56}$Fe, produced in a single CCSN can be estimated from light curve fittings. 
The typical $^{56}$Ni mass per a CCSN is frequently taken to be $\sim 0.1\ M_\odot$. 
\cite{2009MNRAS.395.1409S} reported the observationally determined $^{56}$Ni masses for nearby type IIP SNe with progenitor detections or upper limits on the progenitor's luminosity. 
The derived $^{56}$Ni masses broadly distribute in a range from $0.003\ M_\odot$ to $0.1\ M_\odot$. 
Therefore, the $^{56}$Ni averaged over the entire population of type IIP SNe may be smaller than $0.1\ M_\odot$. 
We first adopt the typical value of $0.1\ M_\odot$ and consider effects of reduced $^{56}$Ni masses later. 

Estimating the oxygen mass ejected in individual SNe is relatively challenging. 
Nevertheless, spectroscopic observations of SN ejecta in the nebular phase combined with theoretical modelings of emission lines can be used to roughly estimate the value. 
\cite{2015MNRAS.448.2482J} compared the luminosities of $\lambda\lambda$6300, 6364 [\ion{O}{1}] lines for 12 type IIP SNe with their spectral synthesis models \citep{2014MNRAS.439.3694J}. 
They concluded that these SNe are consistent with explosions of stars with ZAMS masses smaller than $17M_\odot$, suggesting the corresponding oxygen masses smaller than $\sim1\ M_\odot$. 

These observational constraints are based on a limited number of nearby samples and thus more observational data should be accumulated to further constrain the ejected oxygen and nickel masses. 
However, the current observational constraints (oxygen mass less than $\sim 1\ M_\odot$ and $^{56}$Ni mass less than $\sim 0.1\ M_\odot$ per a single type IIP SNe) are in agreement with theoretical yields for massive stars with ZAMS masses in the range of $9\leq M/M_\odot\leq 17$. 

%%%%%%%%%%%%%%%%%%%%%%%
%%%  Oxygen producer
%%%%%%%%%%%%%%%%%%%%%%%
\section{OXYGEN TO IRON RATIO}\label{sec:missing}
The theoretical CCSNe yields presented in Section \ref{sec:IMF_weighted_yield} suggest that massive stars with ZAMS masses in the range of $20$--$30\ M_\odot$ predominantly supply galaxies with oxygen. 
Thus, limiting the mass range of stars exploding as CCSNe to $9\leq M/M_\odot \leq 17$ leads to significantly inefficient oxygen production. 
In this section, we try to obtain a model-independent requirement for the oxygen production site without detailed chemical enrichment models. 
We assume that oxygen is produced not only by type IIP SNe but a population different from type IIP SNe, and consider requirements on this hypothesized population's event rate and oxygen production efficiency, which are compared with other types of CCSNe. 

The oxygen to iron abundance ratios [O/Fe] of individual stars in the Galaxy are defined in such a way that $[A/B]$ gives the logarithm of the number ratio $N_A/N_B$ of an element A to another B normalized by the solar value $N_{A,\odot}/N_{B,\odot}$,
\begin{equation}
[A/B]=\log_{10}\left(\frac{N_A/N_{A,\odot}}{N_B/N_{B,\odot}}\right).
\end{equation}
The distribution of stars on the [Fe/H]-[O/Fe] plane has been used to investigate relative contributions of different types of SNe to the oxygen and iron enrichment in the Galaxy and their natures \citep[e.g.,][]{1986A&A...154..279M,1995ApJS...98..617T,1995MNRAS.277..945T,1996ApJ...462..266Y,1998ApJ...503L.155K,2001ApJ...558..351M}. 
The oxygen to iron ratios [O/Fe] of metal-poor stars distribute around a mean value of [O/Fe]$\simeq 0.5$ (see Figure \ref{fig:OFe_FeH} below).  
We adopt the mean value of [O/Fe]$_\mathrm{mean}=0.5$ and derive a requirement to reproduce the ratio around the mean value. 

\subsection{Type IIP SNe}
We first consider the contribution from type IIP SNe, which is relatively well constrained theoretically and observationally as we have reviewed in Section \ref{sec:CCSNe_yield}. 
The average oxygen and iron masses $m_\mathrm{O}^\mathrm{IIP}$ and $m_\mathrm{Fe}^\mathrm{IIP}$ ejected by a single type IIP SN are calculated in the following way. 
%by averaging the ejected oxygen mass with the implied mass range for type IIP SNe, $9\leq M/M_\odot \leq 17$.  
%By using the yield provided by \citep{2006ApJ...647..483L} and the Kroupa IMF as the weight, the averaged massed yield $m_\mathrm{O}^\mathrm{IIP}=0.3M_\odot$ and $m_\mathrm{Fe}^\mathrm{IIP}=0.1$. 
By integrating the Kroupa IMF, Equation (\ref{eq:Kroupa}), for the implied mass range for type IIP SNe, $9\leq M/M_\odot\leq 17$, the number of type IIP SNe per $1M_\odot$ of star formation is found to be $N^\mathrm{IIP}=5.6\times10^{-3}$ $M_\odot^{-1}$. 
The oxygen and iron yields, $Y^\mathrm{IIP}_\mathrm{O}$ and $Y^\mathrm{IIP}_\mathrm{Fe}$, for the corresponding mass range are also calculated to be $Y^\mathrm{IIP}_\mathrm{O}=2.0\times 10^{-3}$ and $Y^\mathrm{IIP}_\mathrm{Fe}=6.6\times 10^{-4}$ by using \cite{2006ApJ...647..483L} yields. 
This leads to the ejected masses in a single event, $m^\mathrm{IIP}_\mathrm{O}=Y^\mathrm{IIP}_\mathrm{O}/N^\mathrm{IIP}=0.36M_\odot$ and $m^\mathrm{IIP}_\mathrm{Fe}=Y^\mathrm{IIP}_\mathrm{Fe}/N^\mathrm{IIP}=0.12M_\odot$. 

When the chemical enrichment of the Galaxy has not sufficiently proceeded and not been contributed by SNe Ia, i.e., at low [Fe/H], the oxygen to iron ratio [O/Fe] should reflect their production rates by CCSNe. 
Assuming that the oxygen and iron masses ejected by CCSNe weakly depend on the metallicity (Figure \ref{fig:mass_OFeZ}), the ratio implied for type IIP SNe is 
\begin{eqnarray}
\mathrm{[O/Fe]}_\mathrm{IIP}&=&\log_{10}\left
(\frac{m^\mathrm{IIP}_\mathrm{O}}{m^\mathrm{IIP}_\mathrm{Fe}}
\right)
-\log_{10}\left(\frac{m_\mathrm{O}N_\mathrm{O,\odot}}{m_\mathrm{Fe}N_\mathrm{Fe,\odot}}\right)
\nonumber\\
&=&-0.16,
%-0.157174
\label{eq:OFe_iip}
\end{eqnarray}
where $m_\mathrm{O}$ and $m_\mathrm{Fe}$ are the masses of single oxygen and iron atom averaged over different isotopes and $N_\mathrm{O,\odot}$ and $N_\mathrm{Fe,\odot}$ are the solar values of the relative numbers of oxygen and iron. 
This value is much lower than the mean value, [O/Fe]$_\mathrm{mean}=0.5$, suggesting inefficient oxygen production by type IIP SNe to account for a main production site of oxygen.

\subsection{Oxygen Enrichment by a Missing Population}
%%%%%%%%%%%%%%%%%%%%%%%%%%%%%
\begin{figure*}
\begin{center}
\includegraphics[scale=0.8,bb=0 0 535 446]{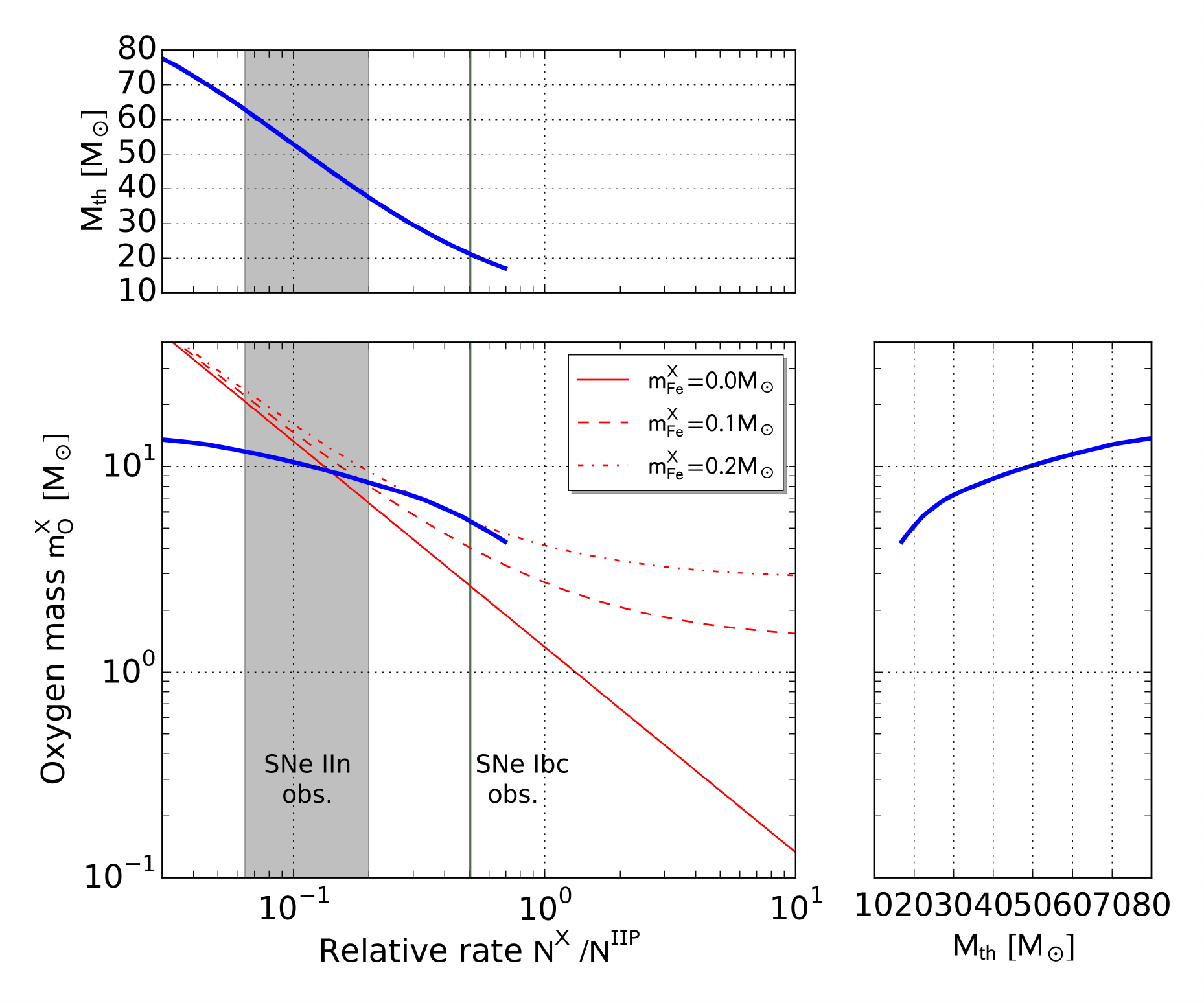}
\caption{Average oxygen mass $m^\mathrm{X}_\mathrm{O}$ and the relative rate $N^\mathrm{x}/N^\mathrm{IIP}$ required to explain the mean oxygen to iron ratio $\mathrm{[O/Fe]}_\mathrm{mean}$ by a combined contribution by type IIP SNe and an unspecified population. 
The thin solid, dashed, and dash-dotted (red) curves in the lower left panel correspond to cases with $m_\mathrm{Fe}^\mathrm{X}/M_\odot=0.0$, $0.1$, and $0.2$. 
The thick (blue) curve in the lower left panel shows the relation between the oxygen mass and relative rate obtained by integrating the IMF from a threshold value $M_\mathrm{th}$ to $M_\mathrm{max}=100\ M_\odot$. 
The oxygen mass and relative rate as functions of the threshold mass are shown in the right and upper panels. 
In the left panels, the ranges corresponding to the relative rates for type Ibc and IIn SNe implied by observations \citep{2009MNRAS.395.1409S,2011MNRAS.412.1522S} are shown as the green vertical line and the gray shaded region. 
}
\label{fig:rate_mass}
\end{center}
\end{figure*}
%%%%%%%%%%%%%%%%%%%%%%%%%%%%%

The significantly small oxygen to iron ratio realized by type IIP SNe clearly indicates that additional contributions are needed. 
Before trying to associate the missing contributor to different types of CCSNe or other phenomena, we consider general requirements for the population. 
We assume that the population produces mass ejection events within a time scale similar to lifetimes of massive stars. 
We denote the rate of the event per $1M_\odot$ of star formation by $N^\mathrm{X}$. 
We further intorduce the oxygen and iron masses produced by a single event, $m^\mathrm{X}_\mathrm{O}$ and $m^\mathrm{X}_\mathrm{Fe}$, in a similar manner to type IIP SNe. 

Since the relative event rate to type IIP SNe is $N^\mathrm{X}/N^\mathrm{IIP}$, the oxygen to iron ratio including the contribution from the population is given by
\begin{eqnarray}
\mathrm{[O/Fe]}_\mathrm{IIP+X}&=&
\log_{10}\left(
\frac{m^\mathrm{IIP}_\mathrm{O}+m^\mathrm{X}_\mathrm{O}N^\mathrm{X}/N^\mathrm{IIP}}
{m^\mathrm{IIP}_\mathrm{Fe}+m^\mathrm{X}_\mathrm{Fe}N^\mathrm{X}/N^\mathrm{IIP}}
\right)
\nonumber\\&&
-\log_{10}\left(\frac{m_\mathrm{O}N_\mathrm{O,\odot}}{m_\mathrm{Fe}N_\mathrm{Fe,\odot}}\right).
%\nonumber
\end{eqnarray}
Therefore, by requiring $\mathrm{[O/Fe]}_\mathrm{IIP+X}=\mathrm{[O/Fe]}_\mathrm{mean}$, we obtain a relation between the three unknown quantities: $m_\mathrm{O}^\mathrm{x}$, $m_\mathrm{Fe}^\mathrm{X}$, and $N^\mathrm{x}/N^\mathrm{IIP}$.

In Figure \ref{fig:rate_mass}, we plot the relation between $m^\mathrm{X}_\mathrm{O}$ and $N^\mathrm{x}/N^\mathrm{IIP}$ for different values of $m^\mathrm{X}_\mathrm{Fe}/M_\odot=0.0$, $0.1$, and $0.2$. 
For example, if this population would have a rate similar to type IIP SNe, $N^\mathrm{X}/N^\mathrm{IIP}\simeq 1$, a single event should produce oxygen masses of $m^\mathrm{X}_\mathrm{O}/M_\odot\simeq 2.5$, $4.0$, and $5.0$ for assumed average iron masses of $m^\mathrm{X}_\mathrm{Fe}/M_\odot=0.0$, $0.1$, and $0.2$. 
The required oxygen mass becomes increasingly larger for lower relative rates. 

\subsection{Theoretical and Observational Constraints}
We compare the requirement for the oxygen mass and the relative rate to type IIP SNe with theoretical yields of CCSNe. 
We assume that the population originates from massive stars with ZAMS masses above a threshold mass $M_\mathrm{th}$ and calculate the relative fraction of the stars to the type IIP progenitors. 
In other words, the relative rate $N^\mathrm{X}/N^\mathrm{IIP}$ is given as follows,
\begin{equation}
\frac{N^\mathrm{X}}{N^\mathrm{IIP}}=\frac{1}{N^\mathrm{IIP}}\int^\mathrm{M_\mathrm{max}}_\mathrm{M_\mathrm{th}}\frac{dN}{dM}dM.
\end{equation}
The average oxygen mass is also expressed as a function of the mass $M_\mathrm{th}$,
\begin{equation}
m^\mathrm{X}_\mathrm{O}=
\frac{1}{N^\mathrm{X}}
\int^\mathrm{M_\mathrm{max}}_\mathrm{M_\mathrm{th}}\frac{dN}{dM}
M_\mathrm{ej}^\mathrm{c}(M,Z)X^\mathrm{cc}_\mathrm{O}(M,Z)dM.
\end{equation}
These quantities are plotted as functions of $M_\mathrm{th}$ in the upper and right panels of Figure \ref{fig:rate_mass}. 
The relation between $m^\mathrm{X}_\mathrm{O}$ and $N^\mathrm{X}/N^\mathrm{IIP}$ is also plotted in the lower left panel of Figure \ref{fig:rate_mass}. 
The resultant curve (thick solid) intersects with the requirements around $N^\mathrm{X}/N^\mathrm{IIP}\simeq 0.1$--$0.7$ and $m^\mathrm{X}_\mathrm{O}\simeq 4$--$8\ M_\odot$, the corresponding mass range of $M_\mathrm{th}\simeq 17$--$40\ M_\odot$. 
Therefore, if stars more massive than $M_\mathrm{th}=17$--$40\ M_\odot$ manage to eject their oxygen layers via some mechanism, then it could account for the observed oxygen to iron ratio [O/Fe]. 

On the other hand, several observations measuring the relative fractions of different types of CCSNe can also constrain their relative rates to type IIP SNe. 
In particular, we focus on the relative rates of type Ibc and IIn SNe, both of which are thought to originate from massive stars having experienced significant mass-loss either by a stellar wind or binary interaction. 
We take into account results reported by \cite{2009MNRAS.395.1409S} and \cite{2011MNRAS.412.1522S}, which are summarized in Table \ref{table:fraction}. 
Based on their volumetric rates, both \cite{2009MNRAS.395.1409S} and \cite{2011MNRAS.412.1522S} found a type Ibc to type IIP fraction of $N^\mathrm{Ibc}/N^\mathrm{IIP}\simeq 0.5$. 
This relative rate is shown as a vertical line in the left panels of Figure \ref{fig:rate_mass}. 
The intersections of the vertical line with the curves shown in the lower left panel of Figure \ref{fig:rate_mass} give the average oxygen mass required to keep the oxygen to iron ratio around the mean value of metal-poor stars. 
For the iron masses ejected by a single Ibc SNe, $m_\mathrm{Fe}^\mathrm{X}/M_\odot=0.0$, $0.1$, and $0.2$, the average oxygen mass is required to be $m^\mathrm{X}_\mathrm{O}/M_\odot\simeq 2.5$, $4$, and $5$. 
A threshold mass of $M_\mathrm{th}=20\ M_\odot$ can account for the oxygen production. 

For type IIn SN rate, systematically different values are reported by \cite{2009MNRAS.395.1409S} and \cite{2011MNRAS.412.1522S}. 
The former study gives $N^\mathrm{IIn}/N^\mathrm{IIP}=3.8/58.7\simeq 0.065$, while the latter gives $N^\mathrm{IIn}/N^\mathrm{IIP}=8.8/42.8\simeq 0.21$. 
Therefore, we show the corresponding range as a shaded area in the left panels of Figure \ref{fig:rate_mass}. 
When we adopt the latter value, $N^\mathrm{IIn}/N^\mathrm{IIP}\simeq 0.21$, the required oxygen mass is $m_\mathrm{O}^\mathrm{X}=6$--$9\ M_\odot$. 
On the other hand, for the former value, $N^\mathrm{IIn}/N^\mathrm{IIP}\simeq 0.065$, the required oxygen mass reaches more than $20\ M_\odot$. 
In fact, as shown in Figure \ref{fig:mass_OFe} and Figure \ref{fig:rate_mass}, producing more than $20\ M_\odot$ in a single massive star is unlikely even for most massive stars. 
Therefore, we conclude that producing sufficient amount of oxygen by type IIn SNe is unlikely when the lower rate for the type IIn SNe is adopted. 
Nevertheless, there is the possibility that the relative rate of type IIn SNe to type IIP SNe is as high as $\sim 0.2$ \citep{2011MNRAS.412.1522S} and that stars more massive than $\sim 40\ M_\odot$ might play a role in the oxygen enrichment. 
%%%%%%%%%%%%%%%%%%%%%%%
%%%  
%%%%%%%%%%%%%%%%%%%%%%%
\section{GALACTIC CHEMICAL EVOLUTION}\label{sec:chemical_evolution}

In this section, we consider the oxygen and iron enrichment in the Galaxy by using a chemical evolution model. 
We have developed a numerical code solving the governing equations of galactic chemical evolution. 
The model adopts the so-called one-box approximation and solves the temporal evolution of the masses of chemical elements in the cold interstellar medium. 
The assumptions and formulations of our model (see Appendix \ref{sec:model} for details) is basically same as simple one-box models presented in various previous work \citep[see, e.g.,][]{1997nceg.book.....P}. 
The dependence of the galactic chemical enrichment history on various assumptions and free parameters has been extensively investigated in the literature \citep[e.g.,][]{2005A&A...430..491R,2010A&A...522A..32R}. 
Although we adopt the IMF in Equation (\ref{eq:Kroupa}) \citep{2001MNRAS.322..231K} in the following calculations, different IMFs can cause systematic differences in chemical abundances from result presented below. 
As we show in Appendix \ref{sec:imf_dependence}, IMFs with a steep gradient in the high-mass range tend to predict 
lower oxygen to iron ratios, which can lead to a systematic downward shift of the [Fe/H]--[O/Fe] relation. 

\subsection{Model Parameters and Assumptions}\label{sec:model_parameters}
We treat two baryonic components in the Galaxy, cold gas and stars. 
We calculate the evolution of the chemical compositions in the two components from $t=0$ to $t=t_\mathrm{max}(=13.8$ Gyr). 
The total masses of the cold gas and star components at $t$ are denoted by $M^\mathrm{g}(t)$ and $M^\mathrm{s}(t)$. 
The masses of chemical elements are $M^\mathrm{g}_i(t)$ and $M^\mathrm{s}_i(t)$ ($i=\mathrm{H},\ \mathrm{He},\cdots$) and these variables evolve according to Equations (\ref{eq:dMg}) and (\ref{eq:dMs}). 

The galactic chemical evolution model needs to specify some free parameters. 
We specify the free parameters so that our fiducial model reproduces observations of Galactic stars. 
The adopted parameters are similar to those found in other recent studies \citep[e.g.,][]{2016ApJ...824...82C,2017ApJ...835..224A}. 
First, the inflow rate $\dot{M}_\mathrm{in}$, at which the cold ISM component increases its mass by accretion, is given by the following exponential form,
\begin{equation}
\dot{M}_\mathrm{in}=\dot{M}_\mathrm{in,0}e^{-t/t_\mathrm{in}},
\end{equation}
where $\dot{M}_\mathrm{in,0}=13.3\ M_\odot\mathrm{yr}^{-1}$ is the normalization constant of the inflow rate and $t_\mathrm{in}=3.0$ Gyr is the time scale characterizing the cold gas accretion. 
The star formation time scale is assumed to be $t_\mathrm{sf}=1$ Gyr. 
The dimensionless parameter $\epsilon_\mathrm{fb}$, which governs the efficiency of the feedback by star formation, is set to be $\epsilon_\mathrm{fb}=2.5$. 

We adopt the CCSNe and AGB yields provided by \cite{2004ApJ...608..405C} and \cite{2010MNRAS.403.1413K}, which cover various ZAMS masses and metallicities. 
Although we consider AGB stars in our calculations, their contributions to the oxygen and iron enrichment in the Galaxy are negligible. 
For type Ia SNe, we adopt the W7 model \citep{1984ApJ...286..644N}, whose ejecta mass and the chemical compositions are provided by \cite{1999ApJS..125..439I}. 
For the iron enrichment, the delay time distribution of type Ia SNe (Equation (\ref{eq:Ia_pow})) is an important input. 
The minimum delay time is assumed to be $t_\mathrm{d,min}=50$ Myr, while the maximum value is identical with the time $t_\mathrm{max}$ when the integration is terminated. 
The delay time distribution is normalized so that the integration of the distribution with respect to time from $t=t_\mathrm{d,min}$ to $t=t_\mathrm{max}$ gives $N_\mathrm{Ia}=1.5\times 10^{-3}$ $M_\odot^{-1}$ \citep{2014ApJ...783...28G,2017ApJ...848...25M}.  
Initially, the cold gas mass is assumed to be $M^\mathrm{g}(0)=10^{7}\ M_\odot$ and the primordial composition is assumed, $X_i^\mathrm{g}=X_{i,0}$,
\begin{equation}
X_{i,0}=\left\{
\begin{array}{ccl}
0.75&\mathrm{for}&i=\mathrm{H},\\ 
0.25&\mathrm{for}&i=\mathrm{He},\\ 
0.0&&\mathrm{otherwise},
\end{array}
\right.
\end{equation}
while the stellar mass is assumed to be zero, $M^\mathrm{s}(0)=0$. 
The mass range of the integration in Equation (\ref{eq:dM_cci}), which gives the mass of element $i$ returning to the cold gas component as CCSN ejecta, should reflect our assumptions on the relation between stars with specific ZAMS mass and their end products, such as different types of CCSNe and compact remnants. 

%%%%%%%%%%%%%%%%%%%%%%%%%%%%%%
\begin{table*}
\begin{center}
  \caption{Chemical evolution models and results}
\begin{tabular}{lccccc}
\hline\hline
Models&mass range [$M_\odot$]&$^{56}$Ni mass [$M_\odot$]
\footnote{We set the nickel masses for $13$ and $15\ M_\odot$ models to these values, while those for the other models ($20$, $25$, $30$, and $35\ M_\odot$) remain unchanged, $M_\mathrm{Ni}=0.1M_\odot$.}&$R_\mathrm{fail}$\footnote{The failed fraction defined by Equation (\ref{eq:R_fail})}&$\log_{10}\epsilon_\mathrm{O}$\footnote{Abundances at the time of the formation of the Sun}&$\log_{10}\epsilon_\mathrm{Fe}$\footnotemark[3]\\
\hline
A10&$9-100$&$0.10$&$0.0$&$8.71$&$7.49$\\
B10&$9-17$&$0.10$&$0.41$&$8.23$&$7.43$\\
C10&$9-17,20-100$&$0.10$&$0.088$&$8.68$&$7.48$\\
D10&$9-17,30-100$&$0.10$&$0.23$&$8.59$&$7.46$\\
E10&$9-17,40-100$&$0.10$&$0.31$&$8.48$&$7.45$\\
A05&$9-100$&$0.05$&$0.0$&$8.71$&$7.46$\\
B05&$9-17$&$0.05$&$0.41$&$8.23$&$7.39$\\
E05&$9-17,40-100$&$0.05$&$0.31$&$8.48$&$7.42$\\
A02&$9-100$&$0.02$&$0.0$&$8.71$&$7.44$\\
B02&$9-17$&$0.02$&$0.41$&$8.21$&$7.37$\\
E02&$9-17,40-100$&$0.02$&$0.31$&$8.48$&$7.40$\\
\hline\hline
\end{tabular}
\label{table:models}
\end{center}
\end{table*}
%%%%%%%%%%%%%%%%%%%%%%%%%%%%%

We carry out calculations with different mass ranges of massive stars producing CCSNe. 
The adopted models and their corresponding mass ranges are listed in Table \ref{table:models}. 
Our fiducial model assumes that all massive stars with $9\leq M/M_\odot\leq 100$ explode as CCSNe and contribute to the metal enrichment (referred to as model A10). 
Next, we restrict the mass range to be $9\leq M/M_\odot\leq 17$ (model B10), which is supposed to produce type IIP SNe. 
As we have done in the previous section, we again consider an additional contribution from stars in upper mass ranges. 
These models assume increasingly larger threshold masses, $20\leq M/M_\odot\leq 100$, $30\leq M/M_\odot\leq 100$, and $40\leq M/M_\odot\leq 100$ (models C10, D10, and E10, respectively), in addition to type IIP SNe. 
These models assume the ejected $^{56}$Ni mass of $0.1\ M_\odot$ in a single event. 

We also investigate effects of reduced $^{56}$Ni mass. 
Observations of type IIP SNe suggest that they produce less $^{56}$Ni than the typically assumed mass of $0.1\ M_\odot$. 
The $^{56}$Ni masses estimated by light curve modelings of most type IIP SNe with reliable constraints on their progenitor masses distribute below the typical value \citep{2009MNRAS.395.1409S}. 
Therefore, reducing the average iron mass produced by stars with ZAMS masses in the range of $9\leq M/M_\odot\leq 17$ may enhance the oxygen to iron ratio. 
\cite{2004ApJ...608..405C} provide CCSNe yields with different values of the mass cut so that users can make their own yield tables with different values of $^{56}$Ni mass via interpolation. 
We reduce the $^{56}$Ni masses produced in models with ZAMS masses of $M=13$ and $15\ M_\odot$ at all metallicities either to $0.05$ or $0.02\ M_\odot$, while keeping those of the remaining models ($M=20,\ 25,\ 30,$ and $35\ M_\odot$) unchanged. 
Since our model determines the mass fractions of elements in the ejected gas from massive stars by those of the nearest stellar mass grid, this treatment effectively reduces the iron mass produced by massive stars with $9\leq M/M_\odot\leq 17.5$. 
We apply this modification to the models A10, B10, and E10. 
The models with the $^{56}$Ni mass of $0.05M_\odot$ ($0.02M_\odot$) are called, A05, B05, and E05 (A02, B02, and E02), respectively. 
The models with reduced $^{56}$Ni masses are also listed in Table \ref{table:models}.

\subsection{Results}

\subsubsection{Temporal Evolution}\label{sec:evolution}
\begin{figure}
\begin{center}
\includegraphics[scale=1.0,bb=0 0 234 390]{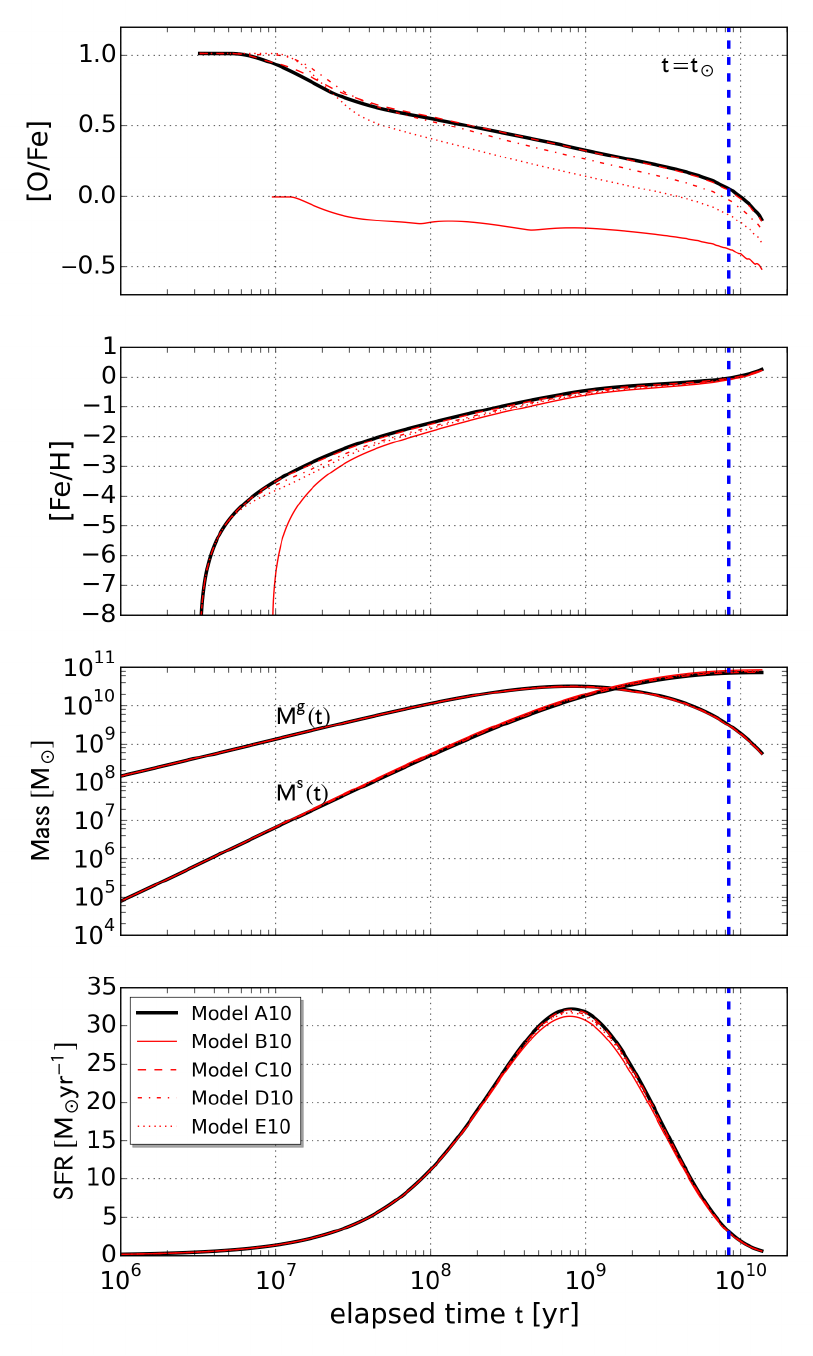}
\caption{Results of the calculations of the galactic chemical evolution. 
The oxygen to iron ratio $[\mathrm{O}/\mathrm{Fe}]$, the iron to hydrogen ratio $[\mathrm{Fe}/\mathrm{H}]$, the cold gas and stellar masses, and the star formation rate are plotted as functions of age $t$ from top to bottom. 
In each panel, the thick solid curve (black) show the model with the ZAMS mass range $9\leq M/M_\odot\leq 100$. 
The thin solid curve (red) corresponds to the model with $9\leq M/M_\odot\leq 17$, while thin dashed, dotted, and dash-dotted curves (red) correspond to the models with the contributions from different upper mass ranges, $20\leq M/M_\odot\leq 100$, $30\leq M/M_\odot\leq 100$, and $40\leq M/M_\odot\leq 100$ in addition to stars with $9\leq M/M_\odot\leq 17$. }
\label{fig:evolution}
\end{center}
\end{figure}

In Figure \ref{fig:evolution}, we show the temporal evolutions of $[\mathrm{O}/\mathrm{Fe}]$ and $[\mathrm{Fe}/\mathrm{H}]$, the cold gas and stellar masses, $M^\mathrm{g}$ and $M^\mathrm{s}$, and the star formation rate $\dot{M}^\mathrm{sf}$ for the models with the different mass ranges. 
The vertical dashed line in each panel represents the time $t_\odot$ of the formation of the Sun, which is defined by $t_\odot=t_\mathrm{max}-4.6$ Gyr.
The solar abundances are supposed to reflect the chemical abundance of the cold gas component at this time. 
The temporal evolutions of the cold gas mass, the stellar mass, and the star formation rate for the different models in the lower two panels of Figure \ref{fig:evolution} do not significantly differ from each other. 
The star formation rate has a peak at around $\sim8\times 10^8$ yr, which is followed by a steady decline to $\sim 0.6\ M_\odot\ \mathrm{yr}^{-1}$ at the end of the calculation $t=13.8$ Gyr.  
The stellar mass continuously increases with time, while the cold gas mass is determined according to the balance between the inflow, the outflow, and the star formation rates. 
The stellar and cold gas masses reach their terminal values of $M^\mathrm{s}(t_\mathrm{max})=8.5\times 10^{10}\ M_\odot$ and $M^\mathrm{g}(t_\mathrm{max})=5.5\times 10^{8}\ M_\odot$ for the model A10. 

%%%%%%%%%%%%%%%%%%%%%%%%%%%%%
%%%%%%%%%%%%%%%%%%%%%%%%%%%%%

Figure \ref{fig:evolution} shows that $[\mathrm{O}/\mathrm{Fe}]$ exhibits significant differences between different models. 
In the model A10, the ratio is initially high ($[\mathrm{O}/\mathrm{Fe}]\simeq 1.0$) due to the contribution from stars more massive than $\sim 20\ M_\odot$. 
These massive stars eject relatively large amounts of oxygen because of their large carbon-oxygen core masses. 
Their short lifetimes ($<8\times 10^6$ yrs for stars more massive than $20\ M_\odot$) realize rapid enrichment of the cold gas by oxygen-rich ejecta. 
The oxygen to iron ratio then decreases with time according to increasing contributions from massive stars in lower mass ranges and type Ia SNe. 
The model B10 shows a significantly small oxygen to iron ratio, $[\mathrm{O}/\mathrm{Fe}]<0.0$, as expected from the consideration in the previous section, Equation (\ref{eq:OFe_iip}). 
In addition, the temporal evolutions of the ratios shown in the upper two panels of Figure \ref{fig:evolution} start at $t\simeq 10^7$ yr. 
This corresponds to the lifetime of the most massive star in the model, $\tau(M=17M_\odot)\simeq 10^7$ yrs. 
In other words, no CCSN occurs until the epoch $t\simeq 10^7$ yr. 
The models C10, D10, and E10, show smaller deviations from the model A10 than the model B10. 
The deviation becomes larger for an increasing threshold mass from $M_\mathrm{th}=20\ M_\odot$ to $M_\mathrm{th}=40\ M_\odot$. 
The iron to hydrogen ratio continuously increases with time and is insensitive to the assumed mass ranges, although the enrichment in the model B10 is delayed from the other models because of the lifetime effect. 
Except for the model B10, the iron to hydrogen ratio increases to [Fe/H]$\simeq -3.5$ before massive stars with $M/M_\odot\leq 20$ starts contributing to the iron enrichment at $t\simeq 10^7$ yr. 
In the model A10, both abundance ratios [O/Fe] and [Fe/H] are close to zero at the time of the formation of the Sun. 
In other words, the solar abundance ratios are reproduced in the model A10.

\subsubsection{Abundance ratios}
\begin{figure*}
\begin{center}
\includegraphics[scale=1.4,bb=0 0 334 223]{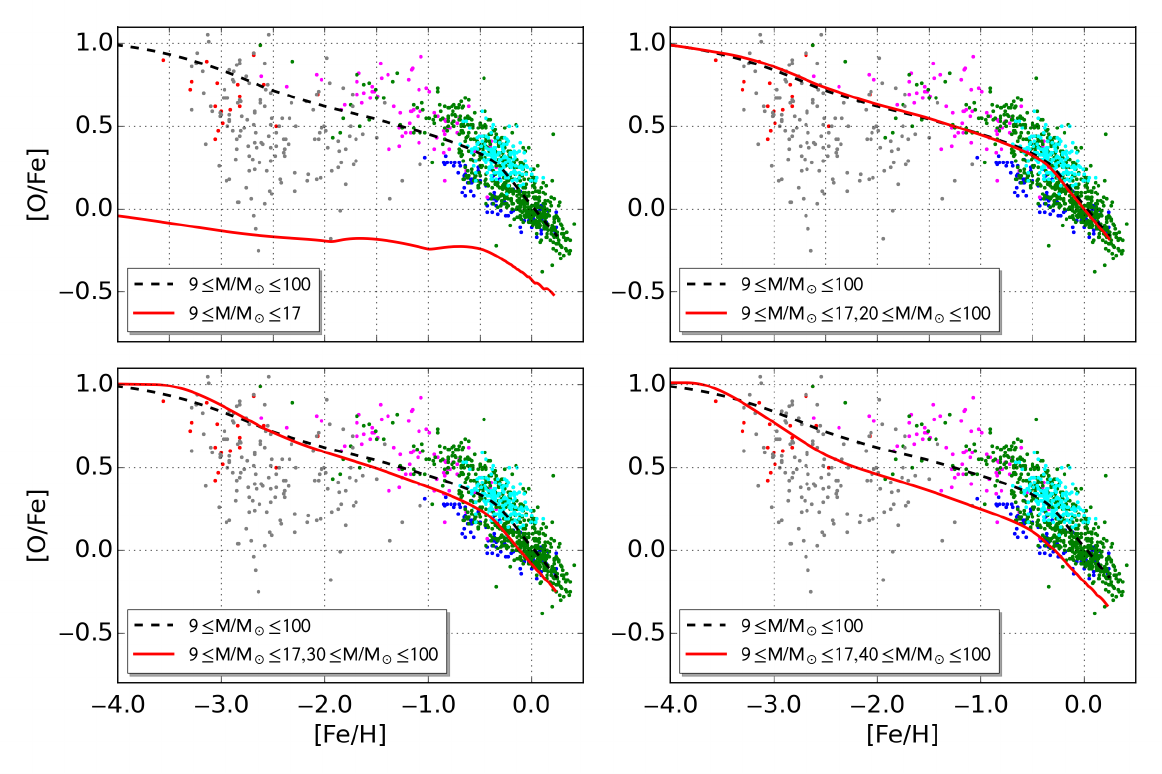}
\caption{Oxygen to iron abundance ratio $[\mathrm{O}/\mathrm{Fe}]$ as a function of the iron to hydrogen ratio $[\mathrm{Fe}/\mathrm{H}]$. 
In each panel, the abundance ratios measured for Galactic stars with various metallicities are plotted as colored dots. 
The data are taken from 
\protect\cite{1993A&A...275..101E} (blue), 
\protect\cite{2003MNRAS.340..304R} (cyan),  
\protect\cite{2003A&A...404..187G} (magenta), 
\protect\cite{2004A&A...416.1117C} (red), 
\protect\cite{2014A&A...562A..71B} (green), 
and \protect\cite{2014AJ....147..136R} (gray). 
The dashed line in each panel shows the result of the galactic chemical evolution model assuming that all massive stars (ZAMS mass range of $9\leq M/M_\odot\leq 100$) explode as CCSNe. 
The solid lines in the four panels show the results obtained by assuming different mass ranges for CCSNe, 
$9\leq M/M_\odot\leq 17$ (upper left), 
$9\leq M/M_\odot\leq 17$ and $20\leq M/M_\odot\leq 100$ (upper right), 
$9\leq M/M_\odot\leq 17$ and $30\leq M/M_\odot\leq 100$ (lower left), 
and $9\leq M/M_\odot\leq 17$ and $40\leq M/M_\odot\leq 100$ (lower right). 
}
\label{fig:OFe_FeH}
\end{center}
\end{figure*}

We compare the temporal evolutions of the abundance ratios obtained by our galactic chemical evolution models with observations of stars in different regions of the Galaxy (solar neighborhood, disk, halo, and so on). 
We have compiled data provided by \cite{1993A&A...275..101E}, \cite{2003MNRAS.340..304R}, \cite{2003A&A...404..187G}, 
\cite{2004A&A...416.1117C}, \cite{2014A&A...562A..71B}, and \cite{2014AJ....147..136R}, who conducted systematic spectroscopic observations of Galactic stars. 
The data are plotted on the [Fe/H]-[O/Fe] planes in Figure \ref{fig:OFe_FeH}. 
We should note that the methods and assumptions employed to obtain elemental abundances, such as local thermodynamics equilibrium (LTE) or non-LTE treatment, can be different from one data set to another, leading to relatively large dispersions on the plot. 
In the panels of Figure \ref{fig:OFe_FeH}, we compare the model B10 (upper left), C10 (upper right), D10 (lower left), and E10 (lower right) with the model A10. 
The model A10, which is our fiducial model, shows a good agreement with the observational trend. 
The high oxygen to iron ratios [O/Fe]$\simeq 1.0$ around [Fe/H]$\simeq -3.0$ are attributed to the metal enrichment by massive stars with $M\geq 20\ M_\odot$ as discussed above. 
The abundance ratios in $-2.0\leq$[Fe/H]$\leq -1.0$ reflect contributions from massive stars in the entire mass range. 
The oxygen to iron ratio [O/Fe]$\simeq 0.5$ in this range is close to the mean value [O/Fe]$_\mathrm{mean}$ adopted in this study. 
Type Ia SNe eventually contribute to the metal enrichment and reduce the oxygen to iron ratio at [Fe/H]$\geq -1$, leading to the widely known break of the trend on the [Fe/H]-[O/Fe] plane.

%%%%%%%%%%%%%%%%%%%%%%%%%%%%%
%%%%%%%%%%%%%%%%%%%%%%%%%%%%%

As we have seen in Section \ref{sec:evolution}, the model B10 fails to reproduce the observed oxygen abundance due to the lack of stars more massive than $17\ M_\odot$. 
This limited mass range makes the oxygen to iron ratio in the model B10 significantly lower than that of the model A10. 
The difference between the models C10 and A10 is not significant and thus the model C10 also successfully explain the observations. 
The model D10 predicts lower [O/Fe] particularly at [Fe/H]$\geq -1.5$. 
Nevertheless, the curve corresponding to the model D10 is still within the dispersion of the observed abundance ratios. 
On the contrary, the model E10 fails to reproduce the observations. 
The analytic consideration in Section \ref{sec:evolution} suggests that the contribution from stars more massive than $40\ M_\odot$ in addition to type IIP progenitors ($9\leq M/M_\odot\leq 17$), which is the same mass range adopted in the model E10, yields the oxygen to iron mass ratio of [O/Fe]$=0.5$. 
In fact, this condition is met in the numerical calculations. 
The oxygen to iron abundance ratio of the model E10 at [Fe/H]$=-2.0$, where the enrichment by massive stars is dominant, agrees with observations, [O/Fe]$_\mathrm{mean}=0.5$. 
However, the subsequent steady decline of the ratio due to type Ia SNe leads to [O/Fe]$\simeq -0.2$ at the time of the formation of the Sun. 

For more quantitative comparisons of the models, we present the oxygen and iron abundances, $\log_{10}\epsilon_\mathrm{O}$ and $\log_{10}\epsilon_\mathrm{Fe}$, at the formation of the sun $t=t_\odot$ in Table \ref{table:models}. 
The abundance for species $A$ is defined as follows,
\begin{equation}
\log_{10}\epsilon_A=\log_{10}(N_A/N_H)+12.
\end{equation}
The fiducial model reproduces the solar oxygen and iron abundances, $\log_{10}\epsilon_\mathrm{O,\odot}=8.69\pm 0.05$ and $\log_{10}\epsilon_\mathrm{Fe,\odot}=7.50\pm 0.04$ \citep{2009ARA&A..47..481A}. 

\subsubsection{Effects of Nickel Mass}

%%%%%%%%%%%%%%%%%%%%%%%%%%%%%
\begin{figure*}
\begin{center}
\includegraphics[scale=1.4,bb=0 0 334 223]{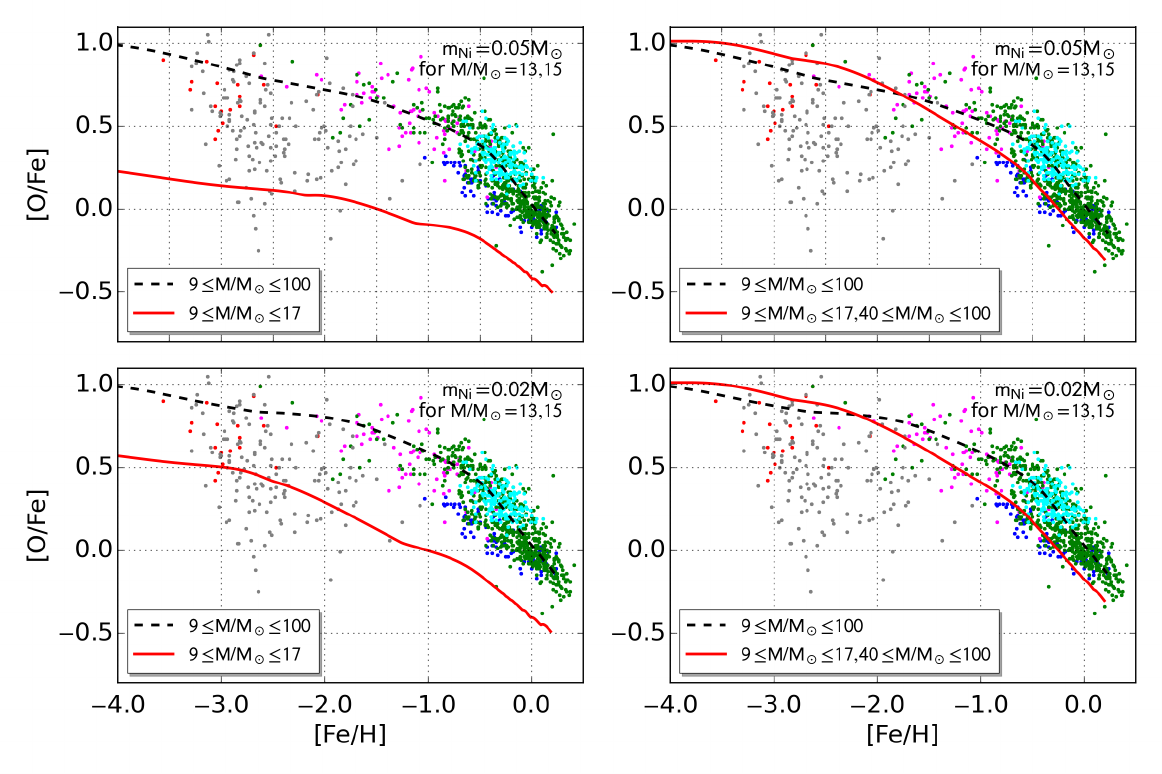}
\caption{Same as Figure \ref{fig:OFe_FeH}, but for models with reduced $^{56}$Ni yields for $13$ and $15\ M_\odot$ stars. 
In the top panels, model B05 (left) and E05 (right) are compared with model A05 (black dashed line in each panel). 
In the lower panels, model B02 (left) and E02 (right) are compared with model A02 (black dashed line in each panel). 
}
\label{fig:OFe_FeH_nickel}
\end{center}
\end{figure*}
%%%%%%%%%%%%%%%%%%%%%%%%%%%%%

One possible way to reconcile the large deviations of the model B10 and E10 from the observed abundance ratios is to reduce the iron mass produced by a single CCSN. 
Figure \ref{fig:OFe_FeH_nickel} shows the models with reduced $^{56}$Ni masses. 
We plot the models with the $^{56}$Ni mass of $0.05\ M_\odot$, the models B05 and E05 in the upper panels, while the lower panels represent the models B02 and E02 with the $^{56}$Ni mass of $0.02\ M_\odot$. 
They are compared with the model A02 and A05, which are represented by black dashed lines in each panel. 
The overall agreement with observations is not obtained even with the reduced $^{56}$Ni production in the lower mass range. 
%The reduction of the $^{56}$Ni mass produced in CCSNe predominantly changes the oxygen to iron ratio at smaller iron to hydrogen ratios [Fe/H]$<-1.5$. 
Even though the oxygen to iron ratio [O/Fe] is improved at [Fe/H]$<-2$, the subsequent enrichment at $[\mathrm{Fe}/\mathrm{H}]>-2$ does not match observations. 
The iron enrichment of the cold gas is delayed due to the reduction. 
As a result, type Ia SNe start contributing the chemical enrichment while the iron enrichment is not sufficient, leading to the break of the curve on the [O/Fe]-[Fe/H] plane at smaller [Fe/H]. 
%Furthermore, since the integrated iron mass produced by CCSNe is small with respect to that by type Ia SNe, the terminal oxygen to iron ratio is below the solar value.
%, indicating that the rate and the delay time distribution of type Ia SN should also be modified. 

\section{DISCUSSION}\label{sec:discussion}

In this section, we consider the final fates of massive stars in the light of the constraints obtained in this work.

\subsection{CCSNe as End Products of Massive Stars}\label{sec:Ibc}
Type IIn and Ibc SNe may potentially compensate the oxygen deficit. 
However, the small relative fraction of type IIn SNe to type IIP SNe,  $N^\mathrm{IIn}/N^\mathrm{IIP}=0.21$ by \cite{2011MNRAS.412.1522S} or $N^\mathrm{IIn}/N^\mathrm{IIP}=0.065$ by \cite{2009MNRAS.395.1409S} indicates that a single type IIn SNe should produce at least more than 8 $M_\odot$ of oxygen, corresponding to stars with ZAMS mass larger than $40\ M_\odot$. 
However, as we have demonstrated by our galactic chemical enrichment calculations, the enrichment of oxygen and iron by CCSNe from stars more massive than $40\ M_\odot$ is not sufficient to reproduce the solar composition. 
On the other hand, type Ibc SNe, whose relative fraction to type IIP SNe is $N^\mathrm{Ibc}/N^\mathrm{IIP}\sim0.5$, require an average oxygen mass of $2.5$--$5\ M_\odot$ per a single event. 
It is worth comparing this constraint with the ejecta mass indicated by observations of type Ibc SNe. 
Recent observational studies of stripped envelope SNe found an average ejecta mass of $\sim 2$--$3\ M_\odot$ and $^{56}$Ni mass of $\sim 0.2\ M_\odot$ for type Ib and Ic SNe \citep{2011ApJ...741...97D,2016MNRAS.457..328L,2017arXiv170707614T}. 
This oxygen mass constraint already exceeds the average ejecta mass implied by observations. 

There are two possibilities reconciling this disagreement. 
First, massive stars with $M\geq$ $20$--$25$ $M_\odot$ indeed explode as type Ibc SNe, but they have shed a part of their oxygen layer prior to the gravitational collapse. 
Since the required oxygen mass for a single event can be as much as $m^\mathrm{X}_\mathrm{O}=5\ M_\odot$, this possibility implies that an oxygen layer with a comparable or larger mass than the ejecta mass inferred by observations should be removed before the iron core-collapse. 
The other possibility is that observed type Ibc SNe originate from stars with ZAMS masses similar to type IIP SNe, which produce a less massive carbon-oxygen core. 
These stars are not expected to completely expel their hydrogen and helium layers via the standard radiatively driven stellar winds. 
Instead, the hydrogen and helium layers should be lost via mass transfer in close binary systems as suggested by several authors \citep[e.g.,][]{2011MNRAS.412.1522S}. 
However, because of less massive oxygen layers developed in those stars, the oxygen production problem remain unsolved in this scenario.

\subsection{Oxygen Layer Ejection and Related Transients}
The former scenario considered above requires that a part of the oxygen layers of massive stars with ZAMS masses larger than $20$--$30$ $M_\odot$ should be ejected in any way. 
Massive stars supposed to predominantly supply oxygen should eject their oxygen layers after their carbon-oxygen cores have grown. 
The theoretical lifetimes of massive stars after having developed their carbon-oxygen core, $\sim 100$--$1000$ yrs \citep[e.g.,][]{2012sse..book.....K}, suggest that the mass-loss rates corresponding to the oxygen layer ejection would be extraordinary, several $10^{-3}$--$10^{-2}$ $M_\odot\ \mathrm{yr}^{-1}$, making this scenario challenging in the standard framework of massive star evolution. 
Furthermore, the final explosion of carbon-oxygen core would occur in the presence of carbon or oxygen-rich circumstellar medium whose mass is comparable to the SN ejecta. 
This could give rise to hypothetical type Ic SNe with narrow emission lines. 
Such SNe have not been observed yet and thus the event rate is expected to be low, making this scenario even more unlikely. 

In the latter scenario, where observed type Ibc SNe originate from less massive stars, the oxygen enrichment would instead be realized by the gravitational collapse of massive stars with $20$--$30\ M_\odot$. 
However, the gravitational collapse should not give rise to bright emission so that the core collapse should not be recognized as any known type of SNe. 
Therefore, this possibility predicts a population of transients less luminous than normal SNe.

\subsection{Theoretical Conditions for CCSNe}
Recently, attempts to obtain criteria judging if a progenitor star could produce a successful CCSN solely from its pre-supernova structure have been paid a great attention. 
The compactness parameter is first introduced by \cite{2011ApJ...730...70O} to quantify the compactness of the central region (a few $M_\odot$ in the mass coordinate) of a collapsing massive star. This parameter is found to well correlate with how difficult the star explode via the neutrino heating mechanism. 
The compactness parameter is a non-monotonic function of the ZAMS mass because it is determined as a consequence of complicated episodes of carbon, oxygen, and silicon burnings in massive stars and mass-losing processes  \citep{2014ApJ...783...10S}. 
It also depends on the numerical treatment of relatively uncertain processes in stellar evolution, such as semi-convection and overshooting. 
The connection between currently available progenitor models and resultant explosions (or non-explosion) has been extensively studied by analytic considerations \citep{2016MNRAS.460..742M} and hydrodynamics simulations with spherical symmetry \citep{2011ApJ...730...70O,2012ApJ...757...69U,2013ApJ...762..126O,2016ApJ...821...38S} and in two dimensions \citep{2015PASJ...67..107N}. 
Despite the wide variety of progenitor models and numerical treatments of the neutrino transfer employed in these studies, they have reached the consensus that massive stars around a ZAMS mass range of $\sim 20$--$30\ M_\odot$ are relatively difficult to make a successful explosion due to their high compactness. 
The compact parameter as a function of the ZAMS mass indeed exhibits a bump around this mass range. 

\cite{2014MNRAS.445L..99H} proposed a critical compactness parameter of $\xi_{2.5}\simeq 0.2$ dividing successful and failed CCSNe. 
This critical value suggests that massive stars with ZAMS masses in the range of $17$--$30\ M_\odot$ collapse without successful CCSNe. 
They argued that this value could also account for the discrepancy between the cosmic star formation rate and the supernova rate implied by observations. 

As we have demonstrated in this study, removing massive stars with ZAMS masses in the range of $17$--$30\ M_\odot$ does not significantly modify the chemical enrichment of oxygen and iron in the Galaxy (see Figure \ref{fig:OFe_FeH}).
Therefore, the theoretical suggestion is in agreement with the non-detection of type IIP SNe with progenitor masses larger than $\simeq 17\ M_\odot$ and the galactic chemical evolution as long as stars more massive than $30\ M_\odot$ successfully eject their oxygen layers. 

\subsection{Failed SNe rate}
Finally we discuss the fraction of stars ending up as failed supernova explosions. 
For a given mass range of massive stars with successful CCSNe, which is denoted by $M_\mathrm{ccsn}$, the fraction of stars leading to iron core-collapses without successful explosions is obtained by integrating the IMF,
\begin{equation}
R_\mathrm{fail}=1-\frac{1}{N_\mathrm{cc}}\int_{M\in M_\mathrm{ccsn}}
\frac{dN}{dM}dM
,
\label{eq:R_fail}
\end{equation}
where $N_\mathrm{cc}$ is the total number of core-collapses per a unit mass,
\begin{equation}
N_\mathrm{cc}=\int^{M_\mathrm{u}}_{M_\mathrm{l}}\frac{dN}{dM}dM\simeq 9.5\times 10^{-3}\ M_\odot^{-1},
\end{equation}
for $M_\mathrm{l}/M_\odot=9$ and $M_\mathrm{u}/M_\odot=100$. 

We present the failed fractions $R_\mathrm{fail}$ for the mass ranges of our models in Table \ref{table:models}. 
The models C10 and D10, in which the oxygen enrichment does not contradict observations, suggest $R_\mathrm{fail}=0.088$ and $0.23$ failed SNe per one core-collapse. 
These numbers agree with the recent result of failed SNe search by the LBT \citep{2017MNRAS.469.1445A}, who suggest the failed fraction of $0.14^{+0.33}_{-0.10}$ at 90\% confidence. 
The failed fraction of the model E10, $R_\mathrm{fail}=0.31$, is still within the error of the failed fraction implied by the observation. 
However, our results of the chemical enrichment calculations suggest a significant deficit of oxygen in the Galaxy. 
The failed fraction of the model B10, $R_\mathrm{fail}=0.41$, is larger than the observational upper limit, leading to a too small number of successful CCSNe. 
Although the error of the failed fraction implied by the observation is still large, future updates would further constrain the failed fraction and tell us if the implied fraction agrees with that from the oxygen production constraint. 

\section{Summary}\label{sec:summary}
In this study, we consider the oxygen and iron enrichment in the Galaxy and the influence of limited mass ranges of massive stars producing CCSNe. 
We have generally considered a population contributing the oxygen and iron enrichment in addition to type IIP SNe and obtained a model-independent requirement for the relative rate of the mass ejection event of the population to type IIP SNe and the average oxygen and iron masses ejected in a single event. 
Only taking into account type IIP SNe, whose mass range and oxygen and iron masses ejected to the surrounding space are relatively well constrained by observations, leads to a significant oxygen deficit as \cite{2013ApJ...769...99B} initially claimed. 
%We have also conducted numerical calculations of galactic chemical enrichment with limited mass ranges. 

In the following, we summarize our findings obtained by our galactic chemical evolution calculations. 
\begin{enumerate}
\item{}First of all, the model assuming that all massive stars with ZAMS masses $9$--$100\ M_\odot$ explode as CCSNe is still most likely in explaining the chemical abundance in solar neighborhood and observations of metal-poor stars in the Galaxy.
\item{}The assumption that massive stars with ZAMS masses $17<M/M_\odot<25$ or $17<M/M_\odot<30$ do not explode as CCSNe does not significantly affect the oxygen and iron enrichment in the Galaxy. 
This mass range is in line with recent theoretical investigations on the explodability of massive stars. 
Removing massive stars in this mass range predicts a failed SN rate of $10$--$20\%$, which is consistent with the recent result of the failed supernova search. 
\item{}
Instead, massive stars with ZAMS masses larger than $30\ M_\odot$ should eject their oxygen layers in some ways to avoid the oxygen deficit problem. 
The oxygen layer ejection by these massive stars should be realized by explosive phenomena other than SNe IIP. 
\item{}
Assuming that massive stars with ZAMS masses larger than $30\ M_\odot$ produce a specific class of transients, its relative rate to type IIP SNe is ~0.3. 
This rate is smaller than the type Ibc SN rate, but is slightly larger than the type IIn SN rate. 
Thus, a fraction of type Ibc SN may be responsible for the oxygen production, if they are contributed sufficiently by explosions of these massive stars. 
However, the ejecta mass distribution of stripped-envelope CCSNe suggested by recent observations is in tension with such massive progenitors. 
\end{enumerate}

\acknowledgments
We appreciate the anonymous referee for his/her constructive comments, which greatly helped us to improve the manuscript. 
This research was supported by MEXT as “Priority Issue on post-K computer” (Elucidation of the Fundamental Laws and Evolution of the Universe) and Joint Institute for Computational Fundamental Science (JICFuS). 
KM acknowledges support by Japan Society for the Promotion of Science (JSPS) KAKENHI Grant 17H02864.

\appendix
\section{GALACTIC CHEMICAL EVOLUTION MODEL}\label{sec:model}
In this section, we introduce the governing equations and parameters of our galactic chemical evolution model. 
Our model assumes the so-called ``one-box'' approximation, where the cold interstellar gas in the galaxy is expected to be uniformly enriched by metals produced by stars. 
The treatment of metal enrichment in the interstellar gas under this approximation has been well established and details of the assumptions and formulations can be found in some textbooks \citep[e.g.,][]{1997nceg.book.....P,2012ceg..book.....M} or papers \citep[e.g.,][]{1980FCPh....5..287T,1986A&A...154..279M,1989MNRAS.239..885M,2006ApJ...653.1145K}. 

\subsection{Governing Equation}
We consider two components, the star and the cold gas, whose masses at age $t$ are denoted by $M^\mathrm{g}(t)$ and $M^\mathrm{s}(t)$. 
Each phase is composed of various chemical elements. 
In our calculations, we consider 31 elements from H to Ga, although we only focus on the oxygen to iron and iron to hydrogen ratios in this study. 
The mass of an element $i$ in the cold gas is denoted by $M^\mathrm{g}_i(t)\ (i=\mathrm{H},\ \mathrm{He},\ \cdots)$. 
Thus, the total mass of the cold gas is obtained by summing up the masses of all the elements considered,
\begin{equation}
M^\mathrm{g}(t)=\sum_iM^\mathrm{g}_i(t). 
\end{equation}
The mass fractions of an element $i$ is given by
\begin{equation}
X^\mathrm{g}_i(t)=\frac{M^\mathrm{g}_i(t)}{M^\mathrm{g}(t)}.
\end{equation}
We define the metallicity of the cold gas in the following way,
\begin{equation}
Z^\mathrm{g}(t)=1-\frac{M^\mathrm{g}_\mathrm{H}(t)+M^\mathrm{g}_\mathrm{He}(t)}{M^\mathrm{g}(t)}.
\end{equation}

These two components {increase/decrease their masses by star formation, feedback, inflow, and mass ejection via stellar winds, SNe Ia, and CCSNe. 
The star formation converts a part of the cold gas to stars at a rate $\dot{M}_\mathrm{sf}$, which is given by the cold gas mass divided by a constant time scale $t_\mathrm{sf}$ characterizing the star formation, $\dot{M}^\mathrm{sf}=M^\mathrm{g}/t_\mathrm{sf}$. 
The cold gas increases its mass by accretion of surrounding gas (referred to as ``inflow'') with the primordial composition. 
The inflow rate $\dot{M}^\mathrm{in}$ is a free parameter. 
Stellar feedback processes, such as SN explosions and stellar winds, bring a part of the cold gas outside the galaxy, depending on the extent of the stellar activity. 
Thus, we model the feedback process by mass loss at a rate proportional to the star formation rate, $\dot{M}^\mathrm{fb}=\epsilon_\mathrm{fb}\dot{M}^\mathrm{sf}$. 
The composition of the material lost from the system is same as that of the cold mass. 
Massive stars with a mass range between $M=M_\mathrm{l}$ and $M=M_\mathrm{u}$ are supposed to end their lives as CCSNe and then return a part of the stellar mass into the cold gas. 
On the other hand, stars less massive than $M=M_\mathrm{l}$ evolve into white dwarfs through AGB stars losing most of their hydrogen and helium layers via AGB winds. 
We denote the rates of these two processes related to stellar activities by $\dot{M}^\mathrm{cc}$ and $\dot{M}^\mathrm{agb}$. 
A fraction of low-mass stars in close binary systems are expected to explode as SNe Ia, whose rate is denoted by $\dot{M}^\mathrm{Ia}$. 
Correspondingly the mass exchange rates for element $i$ are expressed as quantities with subscript $i$, $\dot{M}^\mathrm{cc}_i$, $\dot{M}^\mathrm{agb}_i$, and $\dot{M}^\mathrm{Ia}_i$. 

Using these expressions, the governing equations of the galactic chemical evolution are described as follows,
\begin{equation}
\frac{dM^\mathrm{g}_i}{dt}=\dot{M}^\mathrm{in}_i-\dot{M}^\mathrm{fb}_i-\dot{M}^\mathrm{sf}_i
+\dot{M}^\mathrm{cc}_i+\dot{M}^\mathrm{agb}_i+\dot{M}^\mathrm{Ia}_i,
\label{eq:dMg}
\end{equation}
and
\begin{equation}
\frac{dM^\mathrm{s}_i}{dt}=\dot{M}^\mathrm{sf}_i
-\dot{M}^\mathrm{cc}_i-\dot{M}^\mathrm{agb}_i-\dot{M}^\mathrm{Ia}_i.
\label{eq:dMs}
\end{equation}
We assume that the primordial gas is composed of hydrogen and helium, whose mass fractions are $X_\mathrm{H,0}=0.75$ and $X_\mathrm{He,0}=0.25$. 
Thus, the inflow rate for element $i$ is simply given by $\dot{M}^\mathrm{in}_i=X_{i,0}\dot{M}^\mathrm{in}$. 

\subsection{Mass Ejection From Stars}
\subsubsection{Stellar lifetimes}
When we consider mass ejection from stars associated with their deaths, lifetimes $\tau(M)$ of stars as a function of ZAMS mass $M$ should be taken into account, because finite values of lifetimes lead to delays of mass ejection events with respect to the star formation activity.  
We adopt the following formula presented by \cite{1993ApJ...416...26P},
\begin{equation}
\tau(M)=\left\{
\begin{array}{ccl}
160&\mathrm{Gyr\ for}&0.6\leq M/M_\odot,\\
10^{(0.334-q(M)^{1/2})/0.1116}&\mathrm{Gyr\ for}&0.6<M/M_\odot\leq 6.0,\\
1.2M^{-1.85}+0.003&\mathrm{Gyr\ for}&6.0<M/M_\odot,
\end{array}\right.
\end{equation}
with
\begin{equation}
q(M)=1.790-0.2232\left[7.764-\log_{10}(M)\right].
\end{equation}
(see also \citealt{2005A&A...430..491R}, who studied the dependence of the chemical evolution on uncertainties in stellar lifetimes). 
\subsubsection{CCSNe}
As we have introduced in Section \ref{sec:IMF}, the IMF $dN/dM$ gives the probability distribution of the number of stars with ZAMS mass $M$ when $1M_\odot$ of cold gas is converted to stars. 
The number of stars in a mass range $[M,M+dM]$ supposed to die per unit time at age $t$ is given by the following product, $\dot{M}^\mathrm{sf}[t-\tau(M)](dN/dM)dM$, where the star formation rate is evaluated at $t-\tau(M)$, reflecting the delay of the deaths of the stars from their births. 
Therefore, we obtain the following expression for the mass ejection rate due to CCSNe,
\begin{equation}
\dot{M}^\mathrm{cc}=\int^{M_\mathrm{u}}_{M_\mathrm{l}}M_\mathrm{ej}^\mathrm{cc}(M,Z)\frac{dN}{dM}\dot{M}^\mathrm{sf}[t-\tau(M)]dM,
\label{eq:dM_cc}
\end{equation}
while the rate for element $i$ is given by
\begin{equation}
\dot{M}^\mathrm{cc}_i=\int^{M_\mathrm{u}}_{M_\mathrm{l}}X^\mathrm{cc}_i(M,Z)M_\mathrm{ej}^\mathrm{cc}(M,Z)\frac{dN}{dM}\dot{M}^\mathrm{sf}[t-\tau(M)]dM.
\label{eq:dM_cci}
\end{equation}
As we noted in Section \ref{sec:CCSNe_yield}, the ejecta mass and the mass fraction for stars with various sets of ZAMS mass and metallicity are provided by several groups \cite[e.g.,][]{1995ApJS..101..181W,2004ApJ...608..405C,2006ApJ...647..483L,2006NuPhA.777..424N}.
\subsubsection{AGB winds}
We can derive the following expressions for mass ejection from AGB stars in a similar way to CCSNe,
\begin{equation}
\dot{M}^\mathrm{agb}=\int^{M_\mathrm{l}}_{M_\mathrm{min}}M_\mathrm{ej}^\mathrm{agb}(M,Z)\frac{dN}{dM}\dot{M}^\mathrm{sf}[t-\tau(M)]dM,
\end{equation}
for the total mass and
\begin{equation}
\dot{M}^\mathrm{agb}_i=\int^{M_\mathrm{l}}_{M_\mathrm{min}}X^\mathrm{agb}_i(M,Z)M_\mathrm{ej}^\mathrm{agb}(M,Z)\frac{dN}{dM}\dot{M}^\mathrm{sf}[t-\tau(M)]dM,
\end{equation}
for element $i$, where $M_\mathrm{ej}^\mathrm{agb}(M,Z)$ and $X^\mathrm{agb}_i(M,Z)$ represent the mass ejected as AGB wind and the mass fractions of element $i$. 
We use the values provided by \cite{2010MNRAS.403.1413K} in our calculations. 
\subsubsection{Type Ia SNe}

SNe Ia are expected to occur with a delay to the star-forming activity producing their progenitor system. 
Thus, as in many galactic chemical evolution models, we introduce the delay time distribution of SNe Ia. 
The delay time distribution $D(t_\mathrm{d})$ gives the probability distribution of the number of SNe Ia exploding with a delay time $t_\mathrm{d}$. 
The distribution is normalized so that the following integration gives the number of SNe Ia per unit mass, $N_\mathrm{Ia}$,
\begin{equation}
\int^{t_\mathrm{d,max}}_{t_\mathrm{d,min}}D(t_\mathrm{d})dt_\mathrm{d}=N_\mathrm{Ia},
\end{equation}
where $t_\mathrm{d,min}$ is the minimum delay time, until which no SNe Ia occurs. 
The maximum delay time $t_\mathrm{d,max}$ is simply assumed to be the final age of the system.  
The minimum delay time is assumed to be $50$ Myr to accommodate the prompt population of SNe Ia \citep{2005A&A...433..807M,2006ApJ...648..868S}. 
There are several observational attempts to constrain the functional form and the normalization of the delay time distribution \citep[see,][and reference therein]{2012PASA...29..447M,2014ARA&A..52..107M}. 
In our calculations, we assume the delay time distribution is given by the following power-law form,
\begin{equation}
D(t_\mathrm{d})=
\frac{N_\mathrm{Ia}t^{-1}}{\ln (t_\mathrm{d,max}/t_\mathrm{d,min})}\Theta(t_\mathrm{d}-t_\mathrm{d,min})
\label{eq:Ia_pow}
\end{equation}
with an exponent $-1$, where $\Theta(x)$ is the Heaviside step function. 
Observations suggest the normalization constant of a few $10^{-3}$  SNe $M_\odot^{-1}$ \citep{2012PASA...29..447M,2014ARA&A..52..107M}. 
The normalization constant determines the SNe Ia rate and its ratio to CCSNe rate. 
Several observations suggest the CCSNe to SNe Ia rate of $\sim 3$--$6$ \citep{2009A&A...499..653B,2011MNRAS.412.1441L,2011MNRAS.412.1473L}. 
The iron abundance is also affected by this parameter. 
We adopt $N_\mathrm{Ia}=1.5\times 10^{-3}$ SNe $M_\odot^{-1}$ in our calculations so that our fiducial model reproduces the SNe Ia to CCSNe rate at the end of the calculation and the iron abundance at the time of the formation of the Sun. 
This value is consistent with recent studies \citep[see, e.g., ][]{2014ApJ...783...28G,2017ApJ...848...25M}.

Using the delay time distribution, the contribution of SNe Ia to the increase in the cold gas mass is described as follow,
\begin{equation}
\dot{M}^{\mathrm{Ia}}=
\left\{\begin{array}{ccl}
0&\mathrm{for}&t< t_\mathrm{d,min},
\\
\int^{t-t_\mathrm{d,min}}_{0}M^\mathrm{Ia}_\mathrm{ej}\dot{M}^\mathrm{sf}(t')D(t-t')dt'&\mathrm{for}&t_\mathrm{d,min}\leq t,
\end{array}\right.
\end{equation}
for the total mass and
\begin{equation}
\dot{M}^{\mathrm{Ia}}_i=
\left\{\begin{array}{ccl}
0&\mathrm{for}&t< t_\mathrm{d,min},
\\
\int^{t-t_\mathrm{d,min}}_{0}X_i^\mathrm{Ia}M^\mathrm{Ia}_\mathrm{ej}\dot{M}^\mathrm{sf}(t')D(t-t')dt'&\mathrm{for}&t_\mathrm{d,min}\leq t,
\end{array}\right.
\label{eq:dM_iai}
\end{equation}
for element $i$, where $M_\mathrm{ej}^\mathrm{Ia}$ and $X^\mathrm{Ia}_i$ are the mass and the mass fraction of an element $i$ produced in a single SN Ia. 
We assume the ejecta mass ($M_\mathrm{ej}^\mathrm{Ia}=1.38\ M_\odot$) and the mass fractions realized in the delayed detonation model of SN Ia (W7 model) \citep{1984ApJ...286..644N}, which do not depend on the metallicity. 
The data are taken from \cite{1999ApJS..125..439I}. 

\subsection{Initial mass function}\label{sec:imf_dependence}
The choice of the initial mass function is crucial in chemical enrichment calculations of galaxies. 
Using different IMFs can lead to systematic offsets of chemial abundances from each other. 
In order to evaluate the uncertainties in abundance ratios, we compare our fiducial model with two different calculations adopting IMFs different from Kroupa (2011) IMF. 
We use the IMFs from \cite{1986FCPh...11....1S} and \cite{1993MNRAS.262..545K}.
For the \cite{1986FCPh...11....1S} IMF, we adopt the broken power-law approximation of the IMF \citep{1989MNRAS.239..885M}, which is also adopted by \cite{2005A&A...430..491R}. 
The latter IMF given by \cite{1993MNRAS.262..545K} is originally a broken power-law function. 
The power-law exponents $\alpha$ of these IMFs are given by
\begin{equation}
\alpha=
\left\{
\begin{array}{ccl}
2.35&\mathrm{for}&M/M_\odot<2,\\
2.7&\mathrm{for}&2\leq M/M_\odot,
\end{array}
\right.
\label{eq:Scalo86}
\end{equation}
for \cite{1986FCPh...11....1S} and
\begin{equation}
\alpha=
\left\{
\begin{array}{ccl}
1.3&\mathrm{for}&M/M_\odot<0.5,\\
2.2&\mathrm{for}&0.5\leq M/M_\odot<1,\\
2.7&\mathrm{for}&1\leq M/M_\odot,
\end{array}
\right.
\label{eq:Kroupa93}
\end{equation}
for \cite{1993MNRAS.262..545K}. 
These are referred to as Scalo (1986) IMF and Kroupa (1993) IMF, respectively. 

The most significant difference between the Kroupa (2011) IMF and these two IMFs is the power-law exponent in the high mass regime. 
The former assumes $\alpha=2.3$, while $\alpha=2.7$ for the latter. 
The larger exponent $\alpha$ assumed in Scalo (1986) and Kroupa (1993) IMFs reduces the contribution of massive stars with higher ZAMS masses. 
They especially predict a lower $\alpha$-element to iron ratio, as demonstrated by \cite{2005A&A...430..491R}, who systematically investigated how chemical enrichment calculations depend on assumed IMFs. 
We performed calculations adopting Scalo (1986) and Kroupa (1993) IMFs with the same free parameters as our fiducial model (see Section \ref{sec:model_parameters} for the numerical setup). 
In Table \ref{table:imf}, we present the oxygen and iron abundances obtained by these calculations. 
The oxygen abundances corresponding to the models with the Scalo (1986) and Kroupa (1993) IMFs are smaller by $0.39$ and $0.24$ dex than that of the fiducial model.   
On the other hand, the differences in iron abundances are smaller, because SNe Ia, whose rates in these models are not different from each other, contribute to the iron enrichment almost equally to massive stars. 
In Table \ref{table:imf}, we also provide the CCSNe and SNe Ia rates at the end of the end of the calculation. 
Their values differ from each other only within a factor of 2 and are in agreement with observations. 

%%%%%%%%%%%%%%%%%%%%%%%%%%%%%%
\begin{table}
\begin{center}
  \caption{Dependence on the initial mass function\footnote{The same free parameters as the fiducial model are assumed (see Section \ref{sec:model_parameters}).}}
\begin{tabular}{lrrrr}
\hline\hline
&Kourpa (2011)&Scalo (1986)&Kroupa (1993)\\
\hline
$\log_{10}\epsilon_\mathrm{O}$\footnote{Abundances at the time of the formation of the Sun.}&8.71&8.32&8.47\\
$\log_{10}\epsilon_\mathrm{Fe}$\footnotemark[2]&7.49&7.39&7.44\\
$R_\mathrm{cc}\ [10^{-2}\mathrm{SNe\ yr^{-1}}]$\footnote{The core-collapse supernova rate at the end of the calculation $t=t_\mathrm{max}$.}&1.81&1.26&2.62\\
$R_\mathrm{Ia}\ [10^{-2}\mathrm{SNe\ yr^{-1}}]$\footnote{The type Ia supernova rate at the end of the calculation $t=t_\mathrm{max}$.}&0.426&0.396&0.418\\
$R_\mathrm{cc}/R_\mathrm{Ia}$&4.25&3.19&6.23\\
\hline\hline
\end{tabular}
\label{table:imf}
\end{center}
\end{table}
%%%%%%%%%%%%%%%%%%%%%%%%%%%%%

\end{document}